\documentclass[twocolumn]{aastex61}

\usepackage{color}

\accepted{January 19, 2018}
\submitjournal{AJ}

\begin{document}

\title{HAZMAT III:  The UV Evolution of Mid- to Late-M Stars with {\it GALEX}}

\correspondingauthor{Adam C. Schneider}
\email{aschneid10@gmail.com}

\author{Adam C. Schneider}
\affil{School of Earth and Space Exploration, Arizona State University, Tempe, AZ, 85282, USA}

\author{Evgenya L. Shkolnik}
\affiliation{School of Earth and Space Exploration, Arizona State University, Tempe, AZ, 85282, USA}

\begin{abstract}

Low-mass stars are currently the most promising targets for detecting and characterizing habitable planets in the solar neighborhood. However, the ultraviolet (UV) radiation emitted by such stars can erode and modify planetary atmospheres over time, drastically affecting their habitability.  Thus knowledge of the UV evolution of low-mass stars is critical for interpreting the evolutionary history of any orbiting planets.  \cite{shk14} used photometry from the {\it Galaxy Evolution Explorer (GALEX)} to show how UV emission evolves for early type M stars ($>$0.35 $M_\odot$). In this paper, we extend their work to include both a larger sample of low-mass stars with known ages as well as M stars with lower masses.  We find clear evidence that mid- and late-type M stars (0.08--0.35 $M_\odot$) do not follow the same UV evolutionary trend as early-Ms.  Lower mass M stars retain high levels of UV activity up to field ages, with only a factor of 4 decrease on average in {\it GALEX} NUV and FUV flux density between young ($<$50 Myr) and old ($\sim$5 Gyr) stars, compared to a factor of 11 and 31 for early-Ms in NUV and FUV, respectively.  We also find that the FUV/NUV flux density ratio, which can affect the photochemistry of important planetary biosignatures, is mass and age-dependent for early Ms, but remains relatively constant for the mid- and late-type Ms in our sample.

\end{abstract}


\section{Introduction} \label{sec:intro}

As the field of exoplanet exploration has evolved, low-mass stars, or M stars, have taken a prominent position as objects of interest.  In addition to being the most abundant stellar constituent in the galaxy ($\sim$75\%, \citealt{boch10}) and commonly having terrestrial planets in their habitable zones (HZs) ($\sim$25\%; \citealt{dress15}), M dwarfs have numerous observational advantages for both the identification and characterization of their planets (see e.g., \citealt{shi16}), which have led to several HZ planet discoveries, including a terrestrial planet around our closest stellar neighbor, the M5.5 dwarf Proxima Cen \citep{ang16}, three of seven earth-sized planets around the nearby M8 dwarf TRAPPIST-1 (\citealt{gill16}, \citealt{gill17}), and LHS 1140b, a super-earth around an M4.5 dwarf \citep{ditt17}.

However, because M stars have active chromospheres and coronae that produce high-energy radiation that may be harmful for life, with the added complications due to a prolonged pre-main sequence phase (e.g., \citealt{tian15}) and likely being tidally locked (e.g., \citealt{che17}, \citealt{barnes17}), determining the habitability of planets orbiting M stars is not straightforward.  The stellar UV flux  incident on a planet in an M star's habitable zone may hinder the development extrasolar biology \citep{ranjan17}, destroy the very biosignatures we hope to detect, or even lead to the formation of abiotic oxygen and ozone, producing false-positive biosignatures (\citealt{dom14}, \citealt{tian14}, \citealt{har15}).  Thus knowledge of the UV environments of low-mass stars over planet formation and evolution timescales is vital to any investigation of potentially habitable planets in orbit around them.

Ideally, we would be able to obtain high-resolution UV spectra of a large sample of low-mass stars spanning a wide range of masses and ages.  However, UV spectroscopy is currently only possible with the {\it Hubble Space Telescope}, which is highly competitive and would require more orbits than available for such a sample.  So far, such observations have been limited to a relatively modest sample ($\sim$15) of low-mass stars (\citealt{france13}, \citealt{france16}, \citealt{gui16}, \citealt{young17}).  Alternatively, photometry from the {\it Galaxy Evolution Explorer (GALEX)}, which observed approximately two-thirds of the sky with two UV bands centered at mean wavelengths of 1516 \AA\ (FUV) and 2267 \AA\ (NUV)\footnote{http://galex.stsci.edu/gr6/?page=faq}, provides an invaluable resource for studying the UV evolution and properties of a large sample of low-mass stars.  

There have been several studies utilizing photometry from {\it GALEX} to probe stellar evolution, including using enhanced UV flux to identify new nearby, low-mass, young stars (\citealt{shk11}, \citealt{rod11}, \citealt{rod13}, and \citealt{kast17}), and investigating the age-activity relationship for early type stars (A-K spectral types -- \citealt{find11}).  

\cite{stel13} used a volume limited sample of M stars (10 pc) to investigate correlations between {\it GALEX} and other activity indicators (X-ray and H$\alpha$ emission) and noted a drop in UV flux between young and old M stars.  \cite{ans15} constructed a NUV luminosity function using bright, early-type M dwarfs from \cite{lep11}, and confirmed the results of \cite{stel13} and \cite{shk14} that the base NUV emission for low-mass stars is above that expected from photospheres only.  \cite{jones16} used a sample of M dwarfs from the Palomar/MSU nearby star survey \citep{reid95} and the SDSS DR7 M dwarf spectroscopic sample \citep{west11} to construct a correlation between UV luminosities and H$\alpha$ emission and investigate UV emission versus galactic scale height (a proxy for age), and found a decrease in NUV emission with distance from the plane.  

The Habitable Zones and M dwarf Activity across Time (HAZMAT) program was initiated specifically to determine the time-dependent habitability around low-mass stars by using UV observations to provide the much needed empirical constraints for the short wavelength regions of low-mass stellar models (\citealt{shk14}, \citealt{miles17}).  Planet occurrence rates have been shown to increase with decreasing stellar mass \citep{mul15a}, and the planets around low-mass stars are also typically smaller than those around higher-mass stars \citep{mul15b}.  Considering that the stellar mass function peaks around M4, or $\sim$0.2 $M_\odot$ at old ages \citep{boch10}, mid-Ms may provide the most opportunities and advantageous conditions for detailed characterizations of habitable zone planets through atmospheric spectroscopy.  Several known systems with transiting planets orbiting mid- and late-M dwarfs are already planned to be observed with the {\it James Webb Space Telescope} (e.g., LHS 1140; \citealt{ditt17}, Trappist-1; \citealt{gill16}, \citealt{gill17}), and many more exoplanetary systems around low-mass stars in the Solar neighborhood discovered with the upcoming TESS mission (\citealt{rick14}, \citealt{muir17}, \citealt{stass17}) will be prime candidates for future atmospheric characterization observations (\citealt{bat15}, \citealt{cowan15}, \citealt{cross16}, \citealt{crou17}). In this third installment of the HAZMAT series, we extend the study of the UV environments of low-mass stars at various ages to later spectral types.  We describe our sample selection in Section 2, {\it GALEX} photometry in Section 3, and provide an analysis of the results of this investigation in Section 4.

\section{Low-Mass Star Sample}

We investigated young associations and clusters from \cite{shk14} with confirmed low-mass members.  These included the TW Hya Association (10 $\pm$ 3 Myr; \citealt{bell15}), the $\beta$ Pictoris moving group (24 $\pm$ 3 Myr; \citealt{bell15},  26 $\pm$ 3 Myr; \citealt{niel16}, 22 $\pm$ 6 Myr; \citealt{shk17}), the Tucana-Horologium moving group (45 $\pm$ 4 Myr; \citealt{bell15}), the AB Doradus Moving Group (149$^{+51}_{-19}$; \citealt{bell15}), the Hyades (625 $\pm$ 50 Myr; \citealt{per98})\footnote{Note, however, that \cite{brandt15a} and \cite{brandt15b} determine an age of 750 $\pm$ 100 Myr for the Hyades by including rotation in stellar evolution models.}, and objects with field ages ($\sim$5 Gyr).   

Members of the TW Hya association were taken from the census of stellar and substellar members in \cite{gagne17}.  The objects included in this study are those labeled in \cite{gagne17} as bona fide members and high-likelihood candidate members ($>$90\% probability of belonging to TW Hya).  We chose not to use ``candidate'' TW Hya members from \cite{gagne17}, because, as stated in that paper, this group may suffer from significant contamination from interlopers.  For the $\beta$ Pictoris Moving Group, we use the member list from \cite{shk17} (all ``Y'' and ``Y?'' members from Table 4 of that paper).  Low-mass confirmed members of the Tucana-Horologium Moving Group come from \cite{kraus14}.  There has not yet been a dedicated census of the low-mass members of the AB Doradus Moving Group, so the members come from multiple sources.  References for these sources are provided in Table 5.  Hyades members come from \cite{gold13}, which expanded the census of \cite{roes11} to include more low-mass members using photometry from Pan-STARRS \citep{kais02}.  For field age objects, we use the 8 pc sample of \cite{kirk12} and apply an average age of $\sim$5 Gyr, since there are very few pre-main sequence stars known within 8 pc.      

The same spectral types corresponds to a different masses at different ages.  We determine the corresponding masses for each spectral type for each age by first converting spectral type to effective temperature ($T_{\rm eff}$).  For groups older than 100 Myr, we use Table 5 of \cite{pec13} to find $T_{\rm eff}$ values for each spectral type from M0 to M9.  For the groups younger than 100 Myr, we use Table 6 of \cite{pec13}, which provides adopted $T_{\rm eff}$ values for spectral types of young stars.  Note that the spectral type-$T_{\rm eff}$ relations for young stars in \cite{pec13} only extend to spectral type M5.  For spectral types later than M5 for these young groups, we use the spectral type-$T_{\rm eff}$ relation created for late-M members of the TW Hya association in \cite{gagne17}.  We then interpolate the effective temperatures of group members using the low-mass evolutionary models of \cite{bar15} to estimate masses.  Note that the models of \cite{bar15} have discrete ages; we use the 10 Myr, 25 Myr, 40 Myr, 120 Myr, 625 Myr, and 5 Gyr model grids for TW Hya, $\beta$ Pic, Tuc-Hor, AB Dor, the Hyades, and the field, respectively.  Mass estimates for each spectral type for each age group are given in Table 1.  

In this study, we wish to compare the behavior of mid- and late-type M stars to early-Ms, and chose the fully convective boundary as the division between the two samples.  This limit is taken to be 0.35 $M_{\odot}$ \citep{chab97}, which according to Table 1, occurs between M2 and M3 for all age groups.  We take the hydrogen burning limit (0.08 $M_\odot$; \citealt{burrow97}) as the low-mass cutoff for each age group.  Taking into consideration typical spectral type uncertainties of $\pm$0.5 subtypes, this limit corresponds to a spectral type of $\sim$M5 for TW Hya, $\beta$ Pic, and Tuc-Hor, M6 for AB Dor, and M8 for the Hyades and field age stars.  The mass distribution of all stars used in this study is shown in Figure 1.

\begin{deluxetable*}{ccccccccc}
\tablecaption{Spectral Type-Mass Estimates ($M_\odot$)\tablenotemark{a}}
\tablehead{
Spectral & TW Hya & $\beta$ Pic & Tuc-Hor & AB Dor & Hyades & Field \\
Type & (10 Myr) & (24 Myr) & (45 Myr) & (149 Myr) & (625 Myr) & ($\sim$5 Gyr) }
\startdata
M0 & 0.59 & 0.62 & 0.60 & 0.51 & 0.53 & 0.53 \\
M1 & 0.49 & 0.52 & 0.52 & 0.45 & 0.47 & 0.47 \\
M2 & 0.39 & 0.38 & 0.38 & 0.36 & 0.37 & 0.37 \\
M3 & 0.30 & 0.26 & 0.26 & 0.26 & 0.27 & 0.26 \\
M4 & 0.15 & 0.14 & 0.14 & 0.15 & 0.16 & 0.16 \\
M5 & 0.06 & 0.06 & 0.07 & 0.08 & 0.11 & 0.11 \\
M6 & 0.03 & 0.04 & 0.05 & 0.06 & 0.09 & 0.09 \\
M7 & 0.02 & 0.03 & 0.04 & 0.05 & 0.08 & 0.09 \\
M8 & 0.02 & 0.02 & 0.03 & 0.04 & 0.07 & 0.08 \\
M9 & 0.01 & 0.02 & 0.02 & 0.04 & 0.07 & 0.08 \\
\enddata
\tablenotetext{a}{Based on spectral type-$T_{\rm eff}$ relations from \cite{pec13} and \cite{gagne17} and the models of \cite{bar15}.}
\end{deluxetable*}       

\begin{figure*}
\plotone{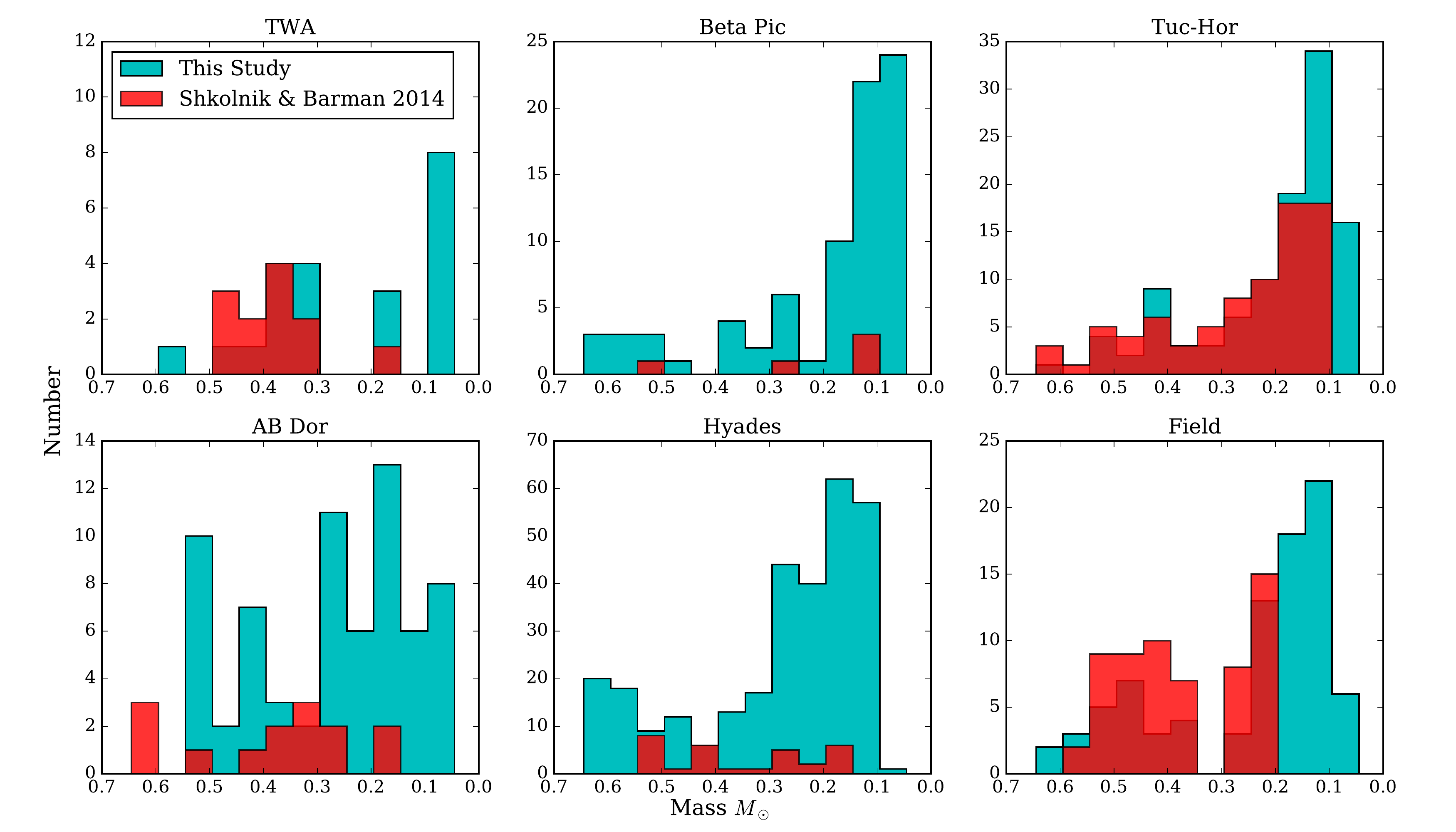}
\caption{The mass distributions of all stellar groups in this study compared to those of \cite{shk14}.  The red histograms represent the \cite{shk14} sample, while the cyan histograms represent all objects in this study observed with the {\it GALEX} NUV passband (i.e., those objects with NUV detections or upper limits).}  
\end{figure*}

\section{{\it GALEX} Photometry}

All photometry gathered from {\it GALEX} for this paper was found using the GalexView tool\footnote{http://galex.stsci.edu/galexview/}.  We used a 10\farcs0 search radius and proper-motion-corrected coordinates (to epoch 2007) for each object.  We exclude all photometry with photometric flags that signal a bright star window reflection, dichroic reflection, detector run proximity, or bright star ghost, as recommended in the {\it GALEX} documentation\footnote{http://www.galex.caltech.edu/wiki/Public:Documentation/Chapter\_103}.  We also exclude all detections with measured magnitudes less than 15 (corresponding to 34 and 108 counts s$^{-1}$ for FUV and NUV, respectively), which marks the threshold for saturation for both the FUV and NUV detectors \citep{mor07}.  A number of objects were found to be either unresolved in the {\it GALEX} images, not observerd by {\it GALEX} at all, or only had {\it GALEX} photometry flagged as unreliable in both the FUV and NUV bands.  The number of objects in each of these categories for each age group is provided in Table 2.  The last column in this table gives the final number of M stars used for each group.

\begin{deluxetable*}{lccccccc}
\tablecaption{Input Sample Summary}
\tablehead{
\colhead{Cluster/Group} & Age & Input Sample & Not Observed & Unresolved & Bad Photometry & Final \\
 & & Members & By {\it GALEX} & in {\it GALEX} & Flags in {\it GALEX} & Number\\
 & (Myr) & (\#) & (\#) & (\#) & (\#) & (\#) }
\startdata
TW Hya & 10 $\pm$ 3 & 40 & 4 & 14 & 0 & 22\\
$\beta$ Pic & 24 $\pm$ 3 & 105 & 0 & 25 & 5 & 75\\
Tuc-Hor & 45 $\pm$ 5 & 118 & 13 & 0 & 0 & 105\\
AB Dor & 149$^{+51}_{-19}$ & 92 & 19 & 7 & 4 & 62\\
Hyades & 625 $\pm$ 50 & 450 & 167 & 2 & 8 & 283\\
Field & $\sim$5000 & 172 & 41 & 35 & 5 & 91\\
\enddata
\end{deluxetable*}

The number of objects observed by {\it GALEX} and detected in the FUV and NUV {\it GALEX} bands is given in Table 3.   The percentage of objects detected was calculated for the entire sample, early-Ms (0.35--0.6 $M_\odot$), and mid- to late-Ms (0.08--0.35 $M_\odot$).  As seen in the table, the fraction of objects detected in the NUV is generally high for each group ($>$65\%).  The fraction of stars detected with the FUV band drops significantly for each group compared to NUV detections, most notably for the Hyades. The drop in detections for Hyades members is likely due to a combination of both distance ($\sim$47 pc; \citealt{van09}) and possibly the older age of the cluster.  We note that the number of detections in most groups is mass dependent.  This is more clearly illustrated in Figure 2, which shows the fraction of stars detected with the NUV and FUV filters as a function of stellar mass for each group.  

\begin{deluxetable*}{lrrrrrrrr}
\tablecaption{{\it GALEX} Detection Summary}
\tablehead{
\colhead{Cluster/Group} & \colhead{FUV} & \colhead{FUV} & \colhead{NUV} & \colhead{NUV} \\
 & (observed) & (detected) & (observed) & (detected) & }
\startdata
\colrule
\sidehead{Entire Sample}
\colrule
TW Hya & 20 & 12 (60\%) & 22 & 20 (91\%) \\
$\beta$ Pic & 67 & 38 (57\%) & 73 & 69 (95\%) \\
Tuc-Hor & 102 & 55 (54\%) & 104 & 96 (92\%) \\
AB Dor & 56 & 41 (73\%) & 61 & 53 (87\%) \\
Hyades & 203 & 28 (14\%) & 280 & 186 (66\%) \\
Field & 78 & 54 (69\%) & 86 & 81 (94\%) \\ 
\colrule
\sidehead{Early-Ms (0.35--0.6 $M_\odot$)}
\colrule
TW Hya & 6 & 6 (100\%) & 7 & 7 (100\%) \\
$\beta$ Pic & 12 & 11 (92\%) & 13 & 13 (100\%) \\
Tuc-Hor & 19 & 18 (95\%) & 19 & 19 (100\%) \\
AB Dor & 21 & 20 (95\%) & 21 & 20 (95\%) \\
Hyades & 49 & 11 (22\%) & 59 & 53 (90\%) \\
Field & 22 & 17 (77\%) & 24 & 23 (96\%) \\ 
\colrule
\sidehead{Mid- to Late-Ms (0.08--0.35 $M_\odot$)}
\colrule
TW Hya & 14 & 6 (43\%) & 15 & 13 (87\%) \\
$\beta$ Pic & 55 & 27 (49\%) & 60 & 56 (93\%) \\
Tuc-Hor & 81 & 37 (46\%) & 85 & 77 (91\%) \\
AB Dor & 35 & 21 (60\%) & 40 & 30 (75\%) \\
Hyades & 154 & 17 (11\%) & 221 & 133 (60\%) \\
Field & 56 & 37 (66\%) & 62 & 58 (94\%) \\ 
\enddata
\end{deluxetable*}

\begin{figure*}
\plotone{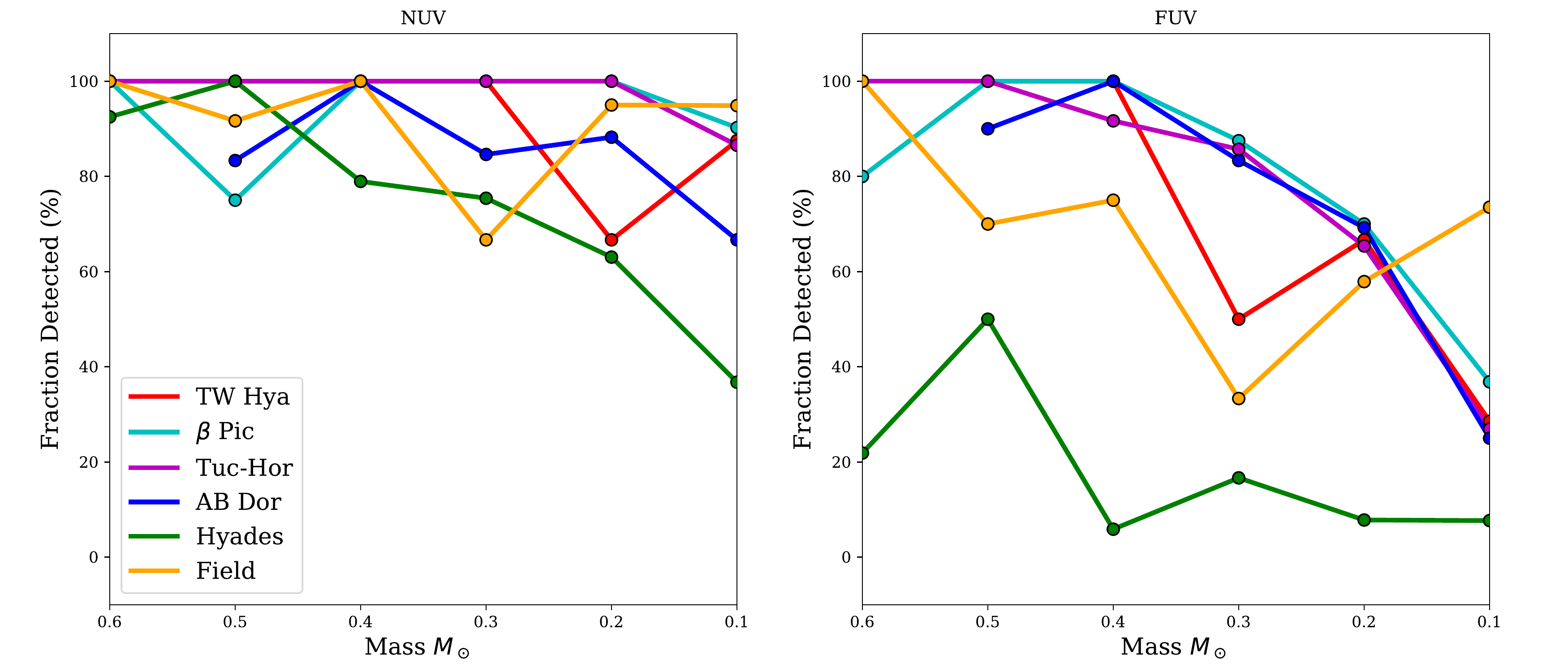}
\caption{The fraction of objects detected with the {\it GALEX} NUV (left) and FUV (right) passbands as a function of mass for all of the groups targeted in this study.}  
\end{figure*}

All {\it GALEX} NUV and FUV photometry is given in Table 5.  In some instances, GalexView will provide multiple detections for a single source, often with varying exposure times.  The magnitudes reported in Table 5 are the weighted averages of all detections and the uncertainties are the weighted standard deviations. 

If a source was observed by {\it GALEX} but not detected, we determine an upper limit for that source by first executing a search around that object's position for all objects within a 10\arcmin\ radius.  We then perform a power-law fit to the relationship between the signal-to-noise ratio (S/N) and the signal of each nearby detection, excluding those objects that exceed the saturation threshold of the detector or with photometric flags listed earlier in this section.  We then take the magnitude that a source would have with a S/N of 2 using our power-law fit as an upper limit.  These upper limits are provided in Table 5.  Note that we do not provide upper limits for sources if they were detected in a different {\it GALEX} exposure.

Figure 1 compares the sample of objects used in this study that were observed with the {\it GALEX} NUV passband to those used for these groups in \cite{shk14}.  \cite{shk14} originally included TWA 31 (M4.2) in their sample of TW Hya members, however this stars has since been shown to be a likely nonmember (\citealt{sch12a}, \citealt{gagne17}) and is not included in this comparison.  As seen in the figure, we have significantly expanded the sample of \cite{shk14}, especially at lower masses.  The expansion of this sample at early types is largely due to concerted efforts to identify more low-mass cluster/group members.  Note that the HAZMAT I study of \cite{shk14} focused mostly on early Ms, and thus included very few known group members with spectral types later than M4.   Since that study, there have been many more investigations targeting and confirming late-M type members of young, nearby moving groups (e.g., \citealt{gagne15}, \citealt{shk17}).

\section{Analysis}

\subsection{The NUV and FUV Evolution of Low-Mass Stars}
To investigate {\it GALEX} FUV and NUV evolutionary trends, we use flux density values relative to the Two Micron All-Sky Survey (2MASS) $J$-band.  We first convert magnitudes to flux densities for {\it GALEX} data in $\mu$Jy using:

\begin{equation}
f_{{\rm GALEX}} = 10^{\frac{23.9 - (\rm m_{GALEX})}{2.5}}
\end{equation}
\\
\noindent where $\rm m_{GALEX}$ is either a {\it GALEX} FUV or NUV magnitude.  To convert 2MASS J magnitudes to flux densities in $\mu$Jy:

\begin{equation}
f_ {\rm J} = (1.594\times10^{9})10^{\frac{\rm m_{J}}{-2.5}}
\end{equation}
\\
\noindent where $\rm m_{J}$ is the 2MASS J-band magnitude.  In order to understand the evolution of the total UV flux density applicable to exoplanet photochemistry models, we do not subtract the contribution from a model photosphere as in \cite{shk14}.  The contribution to FUV and NUV flux densities by the stellar photosphere is discussed in more detail in Section 4.4.

If one wishes to convert flux densities to flux values, care should be taken, specifically with the choice of effective wavelengths for the {\it GALEX} FUV and NUV filters.  Determining the effective wavelength of a filter depends on the underlying spectrum of the object being investigated.   Effective wavelengths are typically determined using the spectrum of Vega, which differs significantly from an M dwarf spectrum (see Appendix for further discussion).
 
Figures 3 and 4 show the distribution of NUV and FUV flux density ratios for each age, split into early-type and mid- to late-type M stars.  As in \cite{shk14}, we see a significant amount of scatter for each group, with ratios typically spanning 1 or 2 orders of magnitude within each age group.  The NUV distributions for early-Ms (left panel of Figure 3) display the same trend as seen in \cite{shk14} and \cite{ans15}, very little change in $f_{\rm NUV}$/$f_{\rm J}$ values up until a decrease is seen at Hyades age, and an even steeper decline from Hyades to field ages.  The distributions of $f_{\rm NUV}$/$f_{\rm J}$ values of lower-mass Ms (right panel of Figure 3) have several notable differences.  First, the $f_{\rm NUV}$/$f_{\rm J}$ ratios for Hyades members are not seen to decrease compared to younger groups.  Secondly, the distribution of $f_{\rm NUV}$/$f_{\rm J}$ values for field objects span a much larger range than that of field early-Ms.  Both of these differences are more clearly seen in Figure 5, which shows $f_{\rm NUV}$/$f_{\rm J}$ ratios as a function of age, and Figure 6, which compares the median $f_{\rm NUV}$/$f_{\rm J}$ values of early-Ms to mid- and late-Ms. 

\begin{figure*}
\plotone{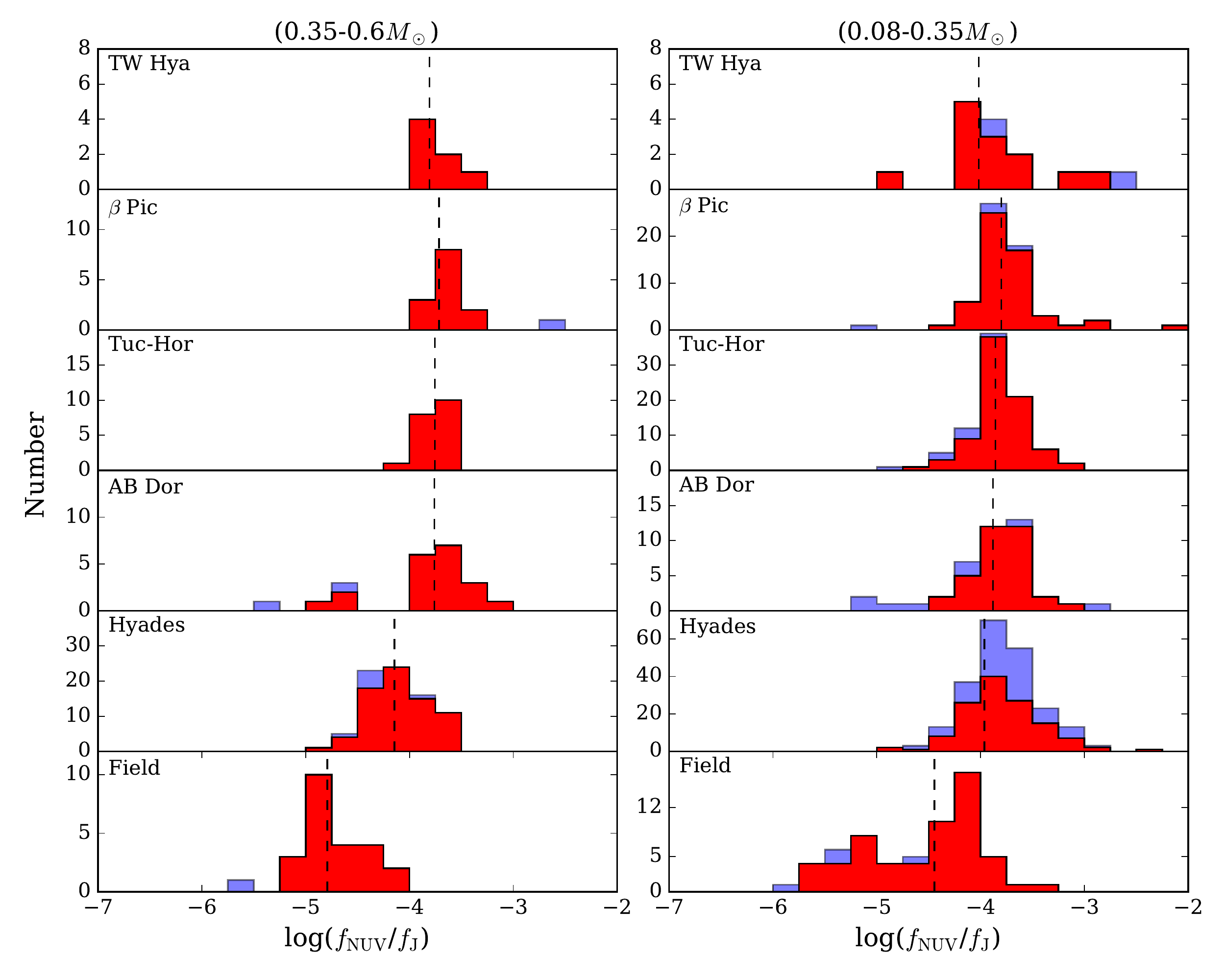}
\caption{Stacked histograms of the {\it GALEX} NUV to 2MASS $J$ flux density ratios for early (left) and mid- to late- (right) objects in this sample.  The red histograms represent detections, while upper limits are shown in blue.  Dashed lines represent the median values given in Table 4.}  
\end{figure*}

\begin{figure*}
\plotone{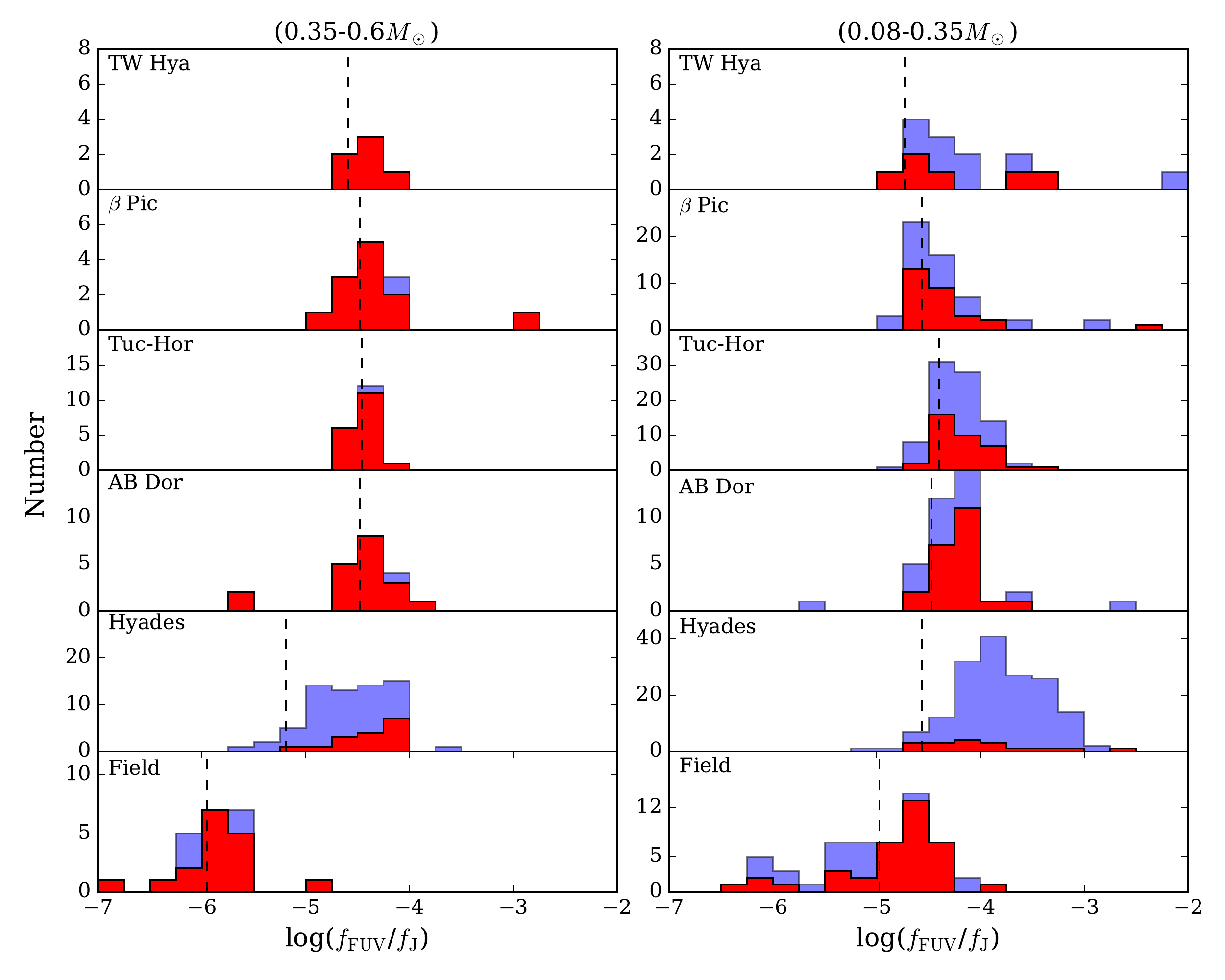}
\caption{Stacked histograms of the {\it GALEX} FUV to 2MASS $J$ flux density ratio for early (left) and mid- to late- (right) objects in this sample.  The red histograms represent detections, while upper limits are shown in blue.  Dashed lines represent the median values given in Table 4.}  
\end{figure*}

\begin{figure*}
\plotone{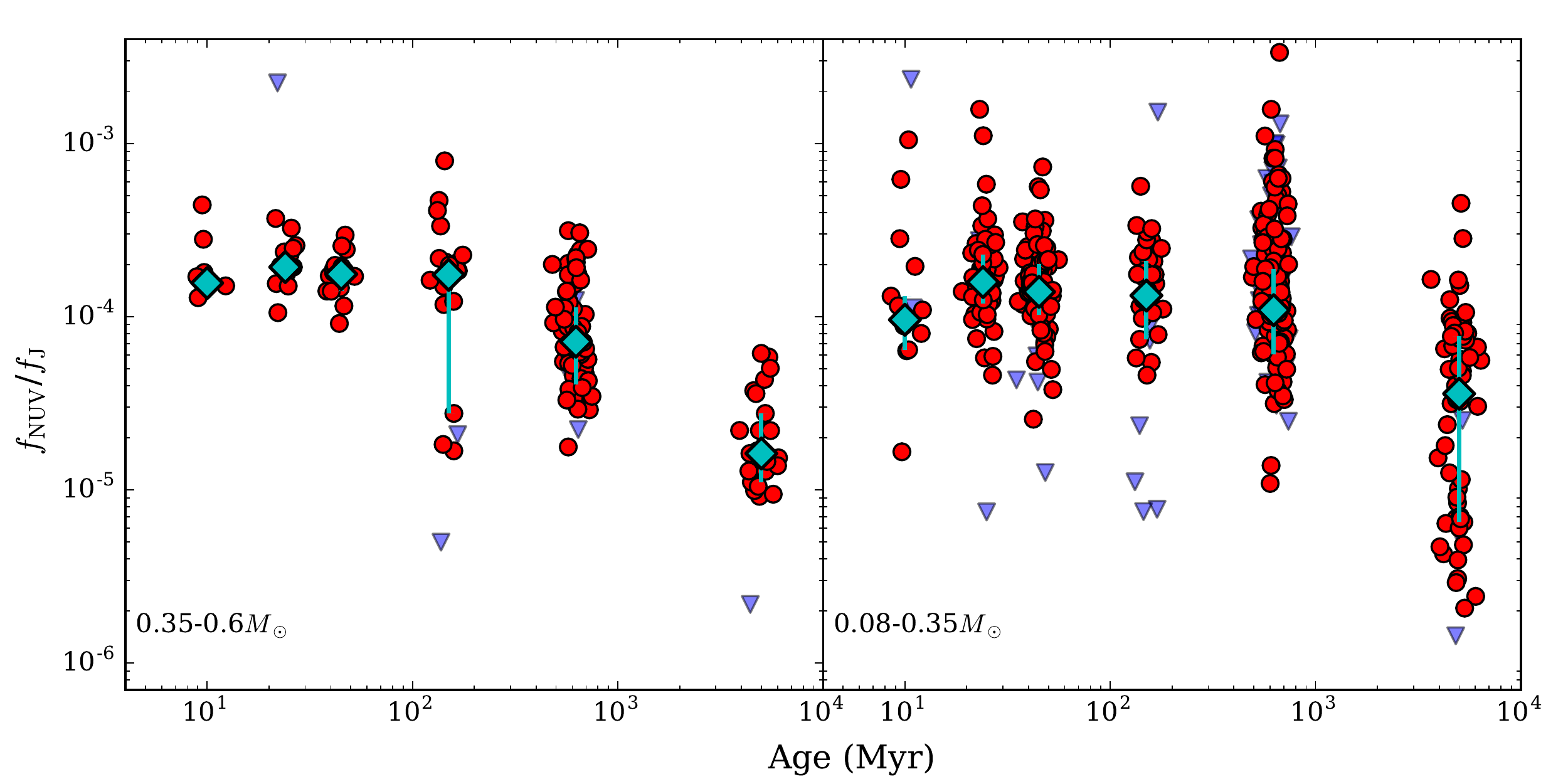}
\caption{{\it GALEX} NUV to 2MASS $J$ fractional flux density as a function of age for all objects in this sample.  Detections are shown as filled red circles, while upper limits are blue triangles. Some symbols are slightly offset along the abscissa for differentiation purposes.  The large cyan diamonds represent the median flux density ratios for each age group.  Error bars on the medians represent inner quartiles as described in Section 4.1. }  
\end{figure*}

\begin{figure}
\plotone{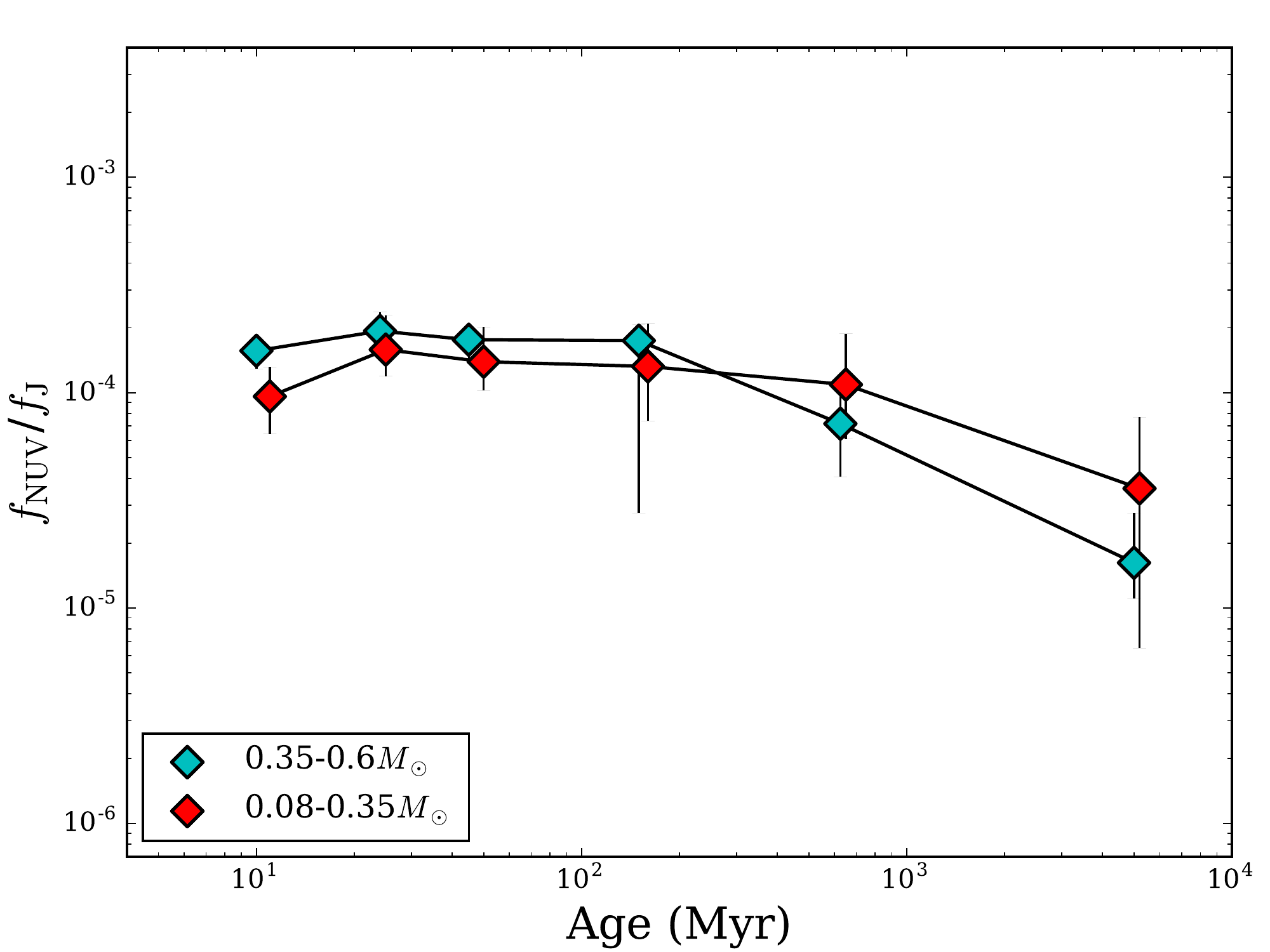}
\caption{The median {\it GALEX} NUV to 2MASS $J$ fractional flux density as a function of age for early-Ms (0.35--0.6 $M_\odot$; cyan) and mid- to late-Ms (0.08--0.35 $M_\odot$; red).  Some symbols are slightly offset along the abscissa for differentiation purposes.    Error bars represent inner quartiles as described in Section 4.1.  }  
\end{figure}

To account for the upper limits in our dataset, we perform a survival analysis using the Kaplan-Meier estimator \citep{kap58} as implemented in the \texttt{lifelines} python package \citep{dav16}.  We treat the lowest $f_{\rm NUV}$/$f_{\rm J}$ values for each group as detections to account for Efron's bias correction \citep{efron67}.  The 50\% values in the resulting empirical cumulative distribution functions are taken as the median values for each group, with 25\% and 75\% values representing the inner quartiles, which are provided in Table 4.  The cumulative distribution functions for early-Ms and mid- to late-Ms for each group are shown in Figure 7.
 
\begin{deluxetable*}{lccccccc}
\tablecaption{Median Fractional Flux Densities}
\tablehead{
\colhead{Cluster/Group} & $f_{\rm FUV}$/$f_{\rm J}$\tablenotemark{a} & $f_{\rm NUV}$/$f_{\rm J}$\tablenotemark{a} }
\startdata
\cutinhead{Early-Ms (0.35--0.6 $M_\odot$)}
TW Hya (10 Myr) & 2.55e-05$_{-2.06{\rm e}-06}^{+1.65{\rm e}-05}$\phn & 1.56e-04$_{-2.77{\rm e}-05}^{+2.35{\rm e}-05}$\phn \\
$\beta$ Pic (24 Myr) & 3.33e-05$_{-8.15{\rm e}-06}^{+4.24{\rm e}-06}$\phn & 1.93e-04$_{-3.80{\rm e}-05}^{+4.32{\rm e}-05}$\phn \\
Tuc-Hor (45 Myr) & 3.49e-05$_{-4.63{\rm e}-06}^{+7.02{\rm e}-06}$\phn & 1.76e-04$_{-3.55{\rm e}-05}^{+1.78{\rm e}-05}$\phn \\
AB Dor (149 Myr) & 3.33e-05$_{-1.22{\rm e}-05}^{+9.46{\rm e}-06}$\phn & 1.74e-04$_{-1.47{\rm e}-04}^{+3.23{\rm e}-05}$\phn \\
Hyades (625 Myr) & 6.49e-06$_{-2.46{\rm e}-06}^{+1.33{\rm e}-05}$\tablenotemark{b} & 7.18e-05$_{-3.11{\rm e}-05}^{+4.18{\rm e}-05}$\phn \\
Field ($\sim$5 Gyr) & 1.13e-06$_{-9.57{\rm e}-07}^{+5.35{\rm e}-07}$\phn & 1.62e-05$_{-5.15{\rm e}-06}^{+1.14{\rm e}-05}$\phn \\
\cutinhead{Late-Ms (0.08--0.35 $M_\odot$)}
TW Hya (10 Myr) & 1.85e-05$_{-1.20{\rm e}-06}^{+3.25{\rm e}-05}$\tablenotemark{b} & 9.60e-05$_{-3.16{\rm e}-05}^{+3.54{\rm e}-05}$\phn \\
$\beta$ Pic (24 Myr) & 2.71e-05$_{-1.28{\rm e}-05}^{+9.20{\rm e}-06}$\phn & 1.58e-04$_{-3.93{\rm e}-05}^{+7.02{\rm e}-05}$\phn \\
Tuc-Hor (45 Myr) & 4.00e-05$_{-2.37{\rm e}-05}^{+1.45{\rm e}-05}$\tablenotemark{b} & 1.39e-04$_{-3.70{\rm e}-05}^{+6.25{\rm e}-05}$\phn \\
AB Dor (149 Myr) & 3.34e-05$_{-8.99{\rm e}-06}^{+4.26{\rm e}-05}$\phn & 1.32e-04$_{-5.84{\rm e}-05}^{+7.74{\rm e}-05}$\phn \\
Hyades (625 Myr) & 2.74e-05$_{-2.09{\rm e}-05}^{+1.70{\rm e}-05}$\tablenotemark{b} & 1.09e-04$_{-4.82{\rm e}-05}^{+7.90{\rm e}-05}$\tablenotemark{b} \\
Field ($\sim$5 Gyr) & 1.06e-05$_{-1.01{\rm e}-05}^{+1.59{\rm e}-05}$\phn & 3.59e-05$_{-2.94e-05}^{+4.10e-05}$\phn \\
\enddata
\tablenotetext{a}{Uncertainties represent inner quartiles of the cumulative distribution functions resulting from the Kaplan-Meier estimator applied to each group.}
\tablenotetext{b}{Note that less than 50\% of observed objects were detected for these subsamples.}
\end{deluxetable*}

\begin{figure*}
\plotone{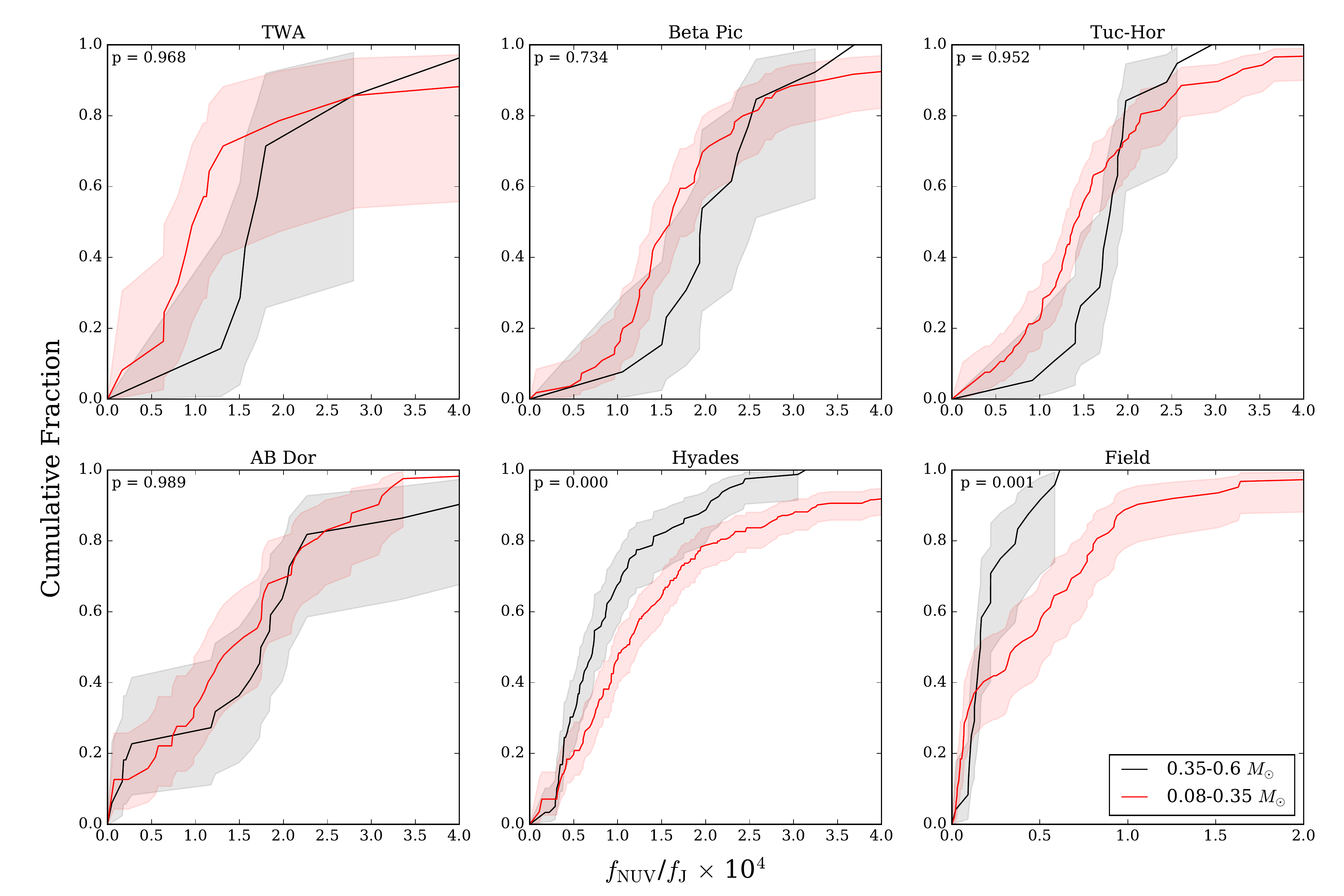}
\caption{{\it GALEX} NUV to 2MASS $J$ fractional flux density cumulative distribution functions for early-Ms (black) and mid- to late-type Ms (red) for all age groups.  The shaded regions represent 1$\sigma$ confidence intervals for each curve.}  
\end{figure*}

These values can be used to predict NUV and FUV flux densities for other low-mass stars, or incident flux densities on exoplanetary atmospheres for hypothetical stars.  This can be accomplished by determining the $J-$band flux density of the object of interest, either through photometry or synthetically using a model spectrum and the $J-$band spectral response curve\footnote{https://www.ipac.caltech.edu/2mass/releases/allsky/doc/sec6\_4a.html}.  Multiplying this $J-$band flux density by the appropriate flux density ratio in Table 4 will give the NUV flux density at the same distance as the $J-$band flux density measurement of the star of interest.   

The $f_{\rm NUV}$/$f_{\rm J}$ values for mid- and late-type field age objects cover a large range of values.  We  investigated whether or not this spread in $f_{\rm NUV}$/$f_{\rm J}$ values was spectral type (or mass) dependent, and found no significant correlation between $f_{\rm NUV}$/$f_{\rm J}$ values and mass for field age mid- and late-type Ms.  We also investigated whether this spread is related to stellar rotation.  We can compare the potential effect of varying rotation rates within our field sample by comparing the rotation period distribution seen for low-mass members of the Hyades \citep{doug16} to that of field M dwarfs seen in \cite{new17}.  \cite{doug16} found that nearly all Hyades stars with masses $\lesssim$0.3 $M_{\odot}$ are rapidly rotating, with over 80\% of such stars having periods less than 10 days and none having periods longer than 30 days.  In contrast, \cite{new17} gathered stellar rotation periods for a sample of nearby M stars from the MEarth Project \citep{berta12}, the majority of which should have ages consistent with the field population, and found a much larger spread in rotation periods for stars with masses less than 0.35 $M_{\odot}$ (see Figure 3 of that paper). Of the stars in the \cite{new17} sample with masses less than 0.35 $M_{\odot}$, $\sim$60\% have rotation periods less than 10 days while $\sim$33\% have periods greater than 30 days.  Thus, at the age of the Hyades, very few mid- to late-Ms have spun down while a significant fraction of mid- to late-Ms have spun down at field ages.  However, there are still a substantial amount of late-M rapid rotators at field ages, with $\sim$33\% having periods less than 1 day.  We therefore conclude that the large range of rotation periods measured for mid- to late-Ms in the sample of \cite{new17} is a plausible explanation for the large range of $f_{\rm NUV}$/$f_{\rm J}$ values for our field age mid- to late-M sample.  This effect is not readily apparent in our early-M sample because the majority early-Ms have spun down at field ages, with $>$60\% of stars with masses between 0.35 and 0.6 $M_{\odot}$ in the \cite{new17} sample having rotation periods greater than 10 days.   

Lastly, the median $f_{\rm NUV}$/$f_{\rm J}$ values for field age mid- to late-Ms is significantly higher than that of early-Ms.  To compare the survival curves of the early and mid- to late-M samples, we use a log-rank parametric test, which accounts for censored data and tests the null hypothesis that two cumulative distributions have the same parent distribution.  The resulting p-values represent the probability that the two populations being compared are drawn from a single distribution.  Thus, a low p-value ($<$0.05) indicates that there is a $>$95\% probability that the differences between the population survival curves are not due to random chance.  We find that the distributions of $f_{\rm NUV}$/$f_{\rm J}$ values for mid- to late-M stars in TW Hya, $\beta$ Pic, Tuc-Hor, and AB Dor are statistically consistent with the $f_{\rm NUV}$/$f_{\rm J}$ values for early-M dwarfs in those groups.  For the Hyades and field-age Ms, we find p-values of 0.000 and 0.001, respectively, indicating a significant difference in the distributions of the early-M and mid- to late-M samples.  This difference can be seen in Figure 7, where the resulting cumulative distribution functions for these groups are distinct. We find that the median $f_{\rm NUV}$/$f_{\rm J}$ ratio decreases by a factor of $\sim$4 from Tuc-Hor age ($\sim$45 Myr) to field ages for mid- to late-Ms, compared to a decrease by a factor of $\sim$11 for early-Ms.  

For the FUV wavelength region, the evolution of the emission from low-mass stars is similar to that of the NUV, with a few key differences.  High-mass Ms in Figure 4 have $f_{\rm FUV}$/$f_{\rm J}$ levels that remain relatively consistent until at least the age of AB Dor (149 Myr), with a sharp drop off as field ages are reached, whereas the $f_{\rm FUV}$/$f_{\rm J}$ values of lower-mass Ms remain relatively constant until  Hyades age (625 Myr), with a much smaller drop off toward field ages. Note, however, that the number of detected objects in the FUV in these groups is significantly less than for the NUV, especially for the Hyades (see Table 3).  

There is a substantial difference between $f_{\rm FUV}$/$f_{\rm J}$ values of early and mid- to late-type field M stars (see Figures 8 and 9).  The median $f_{\rm FUV}$/$f_{\rm J}$ for early-Ms decreases by a factor of $\sim$31 from Tuc-Hor age to field ages, compared to a factor of $\sim$4 for mid- to late-Ms.  Thus, both NUV and FUV flux density measurements indicate that mid- to late-Ms continue to remain more active with increased UV flux density levels up until field ages compared to early-Ms.  A log-rank test comparing the $f_{\rm FUV}$/$f_{\rm J}$ distributions of early Ms and mid- to late-Ms again shows that the distributions for young members of TW Hya, $\beta$ Pic, Tuc-Hor, and AB Dor are not statistically different, while the p-values for the Hyades and field Ms are equal to zero, indicating that these three groups have significantly different samples.
 
\begin{figure*}
\plotone{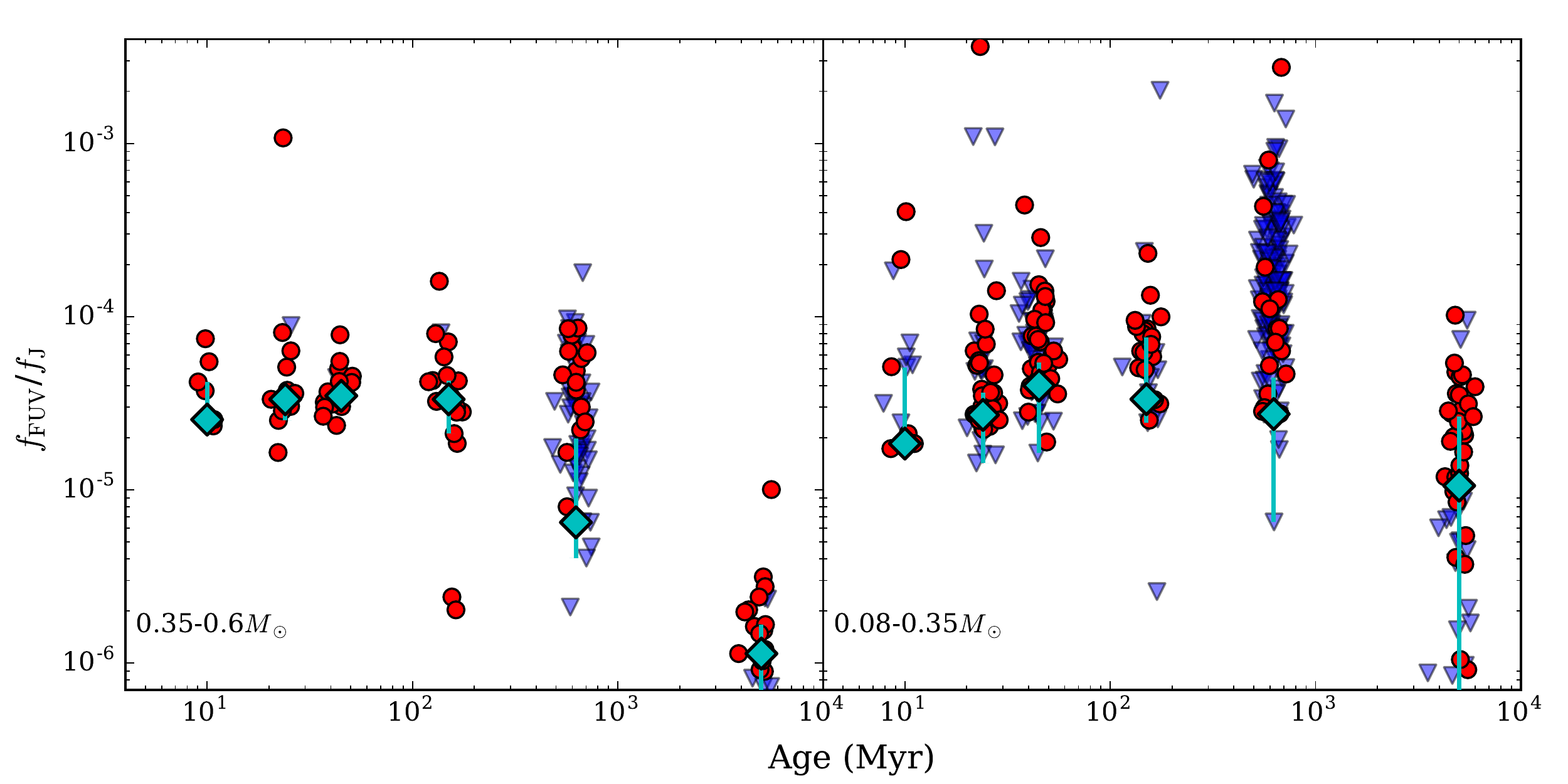}
\caption{{\it GALEX} FUV to 2MASS $J$ fractional flux density as a function of age for all objects in this sample.  Detections are shown as filled red circles, while upper limits are blue triangles.  Some symbols are slightly offset along the abscissa for differentiation purposes.  The large cyan diamonds represent the median flux density ratios for each age group.  Error bars on the medians represent inner quartiles as described in Section 4.1.}  
\end{figure*}

\begin{figure}
\plotone{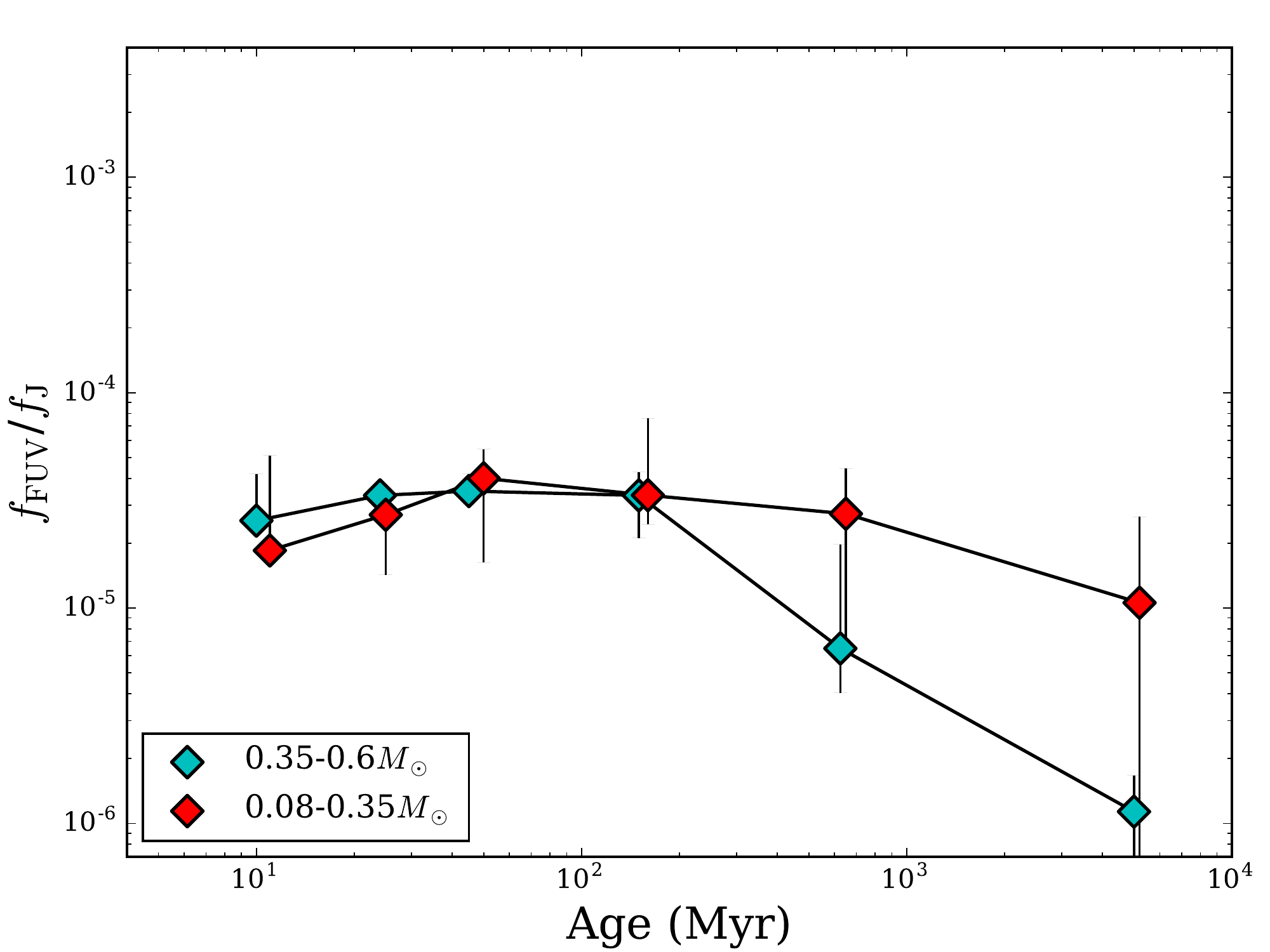}
\caption{The median {\it GALEX} FUV to 2MASS $J$ fractional flux density as a function of age for (0.35--0.6 $M_\odot$; cyan) and mid- to late-Ms (0.08--0.35 $M_\odot$; red).  Some symbols are slightly offset along the abscissa for differentiation purposes.  Error bars represent inner quartiles as described in Section 4.1.  }  
\end{figure}

There have been numerous studies of the activity-rotation relationships for low-mass stars showing that more rapidly rotating stars also show higher levels of activity (e.g., \citealt{delf98}, \citealt{moh03}, \citealt{pizz03}, \citealt{kir07}, \citealt{rein08}, \citealt{brown10}, \citealt{kraus11}, \citealt{rein12}, \citealt{west15}, \citealt{hou16}, \citealt{stel16}, \citealt{ast17}, \citealt{hou17}, \citealt{new17}).  Recent studies of the Hyades \citep{doug16} and the similarly-aged Praesepe cluster \citep{doug17} have shown that M stars with spectral types later than $\sim$M3 continue to rotate very rapidly until at least the ages of these clusters.  If the mechanism that causes M stars to spin down as they age, typically ascribed to a magnetized stellar wind, does not function efficiently for fully-convective stars, then we would expect them to remain in states of rapid rotation for much longer portions of their lifetimes than early-Ms.  Because activity is directly related to rotation, this implies that mid- to late-Ms should remain active for much of their lifetimes.  Thus, the rotation rates of low-mass stars is likely responsible for the {\it GALEX} UV evolutionary trends that we see.

\subsection{Binarity}
Another property that can increase activity levels in low-mass stars is binarity (e.g., \citealt{morgan12}).  A recent investigation of low-mass stellar activity and rotation using data from K2 showed a clear distinction between activity levels of fast and slow rotators, where those stars with short rotation periods are more active \citep{stel16}.  They also showed that a substantial number ($>$40\%) of these fast rotators are visual binaries, compared to $\sim$3\% for slow rotators.  The increased activity levels seen in binaries when compared to single stars is thought to be due to spinning up through tidal interactions or by a reduction in angular momentum loss \citep{stel16}.  Indeed, evidence that binarity affects stellar rotation rates has been seen in several nearby open clusters, including the Pleiades \citep{covey16}, the Hyades \citep{doug16}, and Praesepe \citep{doug17}.  For this reason, all known visual binaries (VB) and spectroscopic binaries (SB) are flagged in Table 5, with references provided.  

In order to investigate whether or not binarity may be biasing any of our results, we split our sample into stars without any evidence of binarity and known spectroscopic binaries or visual binaries with separations smaller than 5\arcsec, the FWHM of the {\it GALEX} NUV band \citep{mor07}.  Figure 10 shows a comparison of the flux density ratios of these samples for every object detected in the NUV.  We perform a log-rank test to compare the distributions of binaries versus the rest of each sample and find p-values $>$0.2 for all groups, consistent with both groups being drawn from the same population.  Thus the inclusion of binaries in our evolutionary studies does not bias our results.   

\begin{figure*}
\plotone{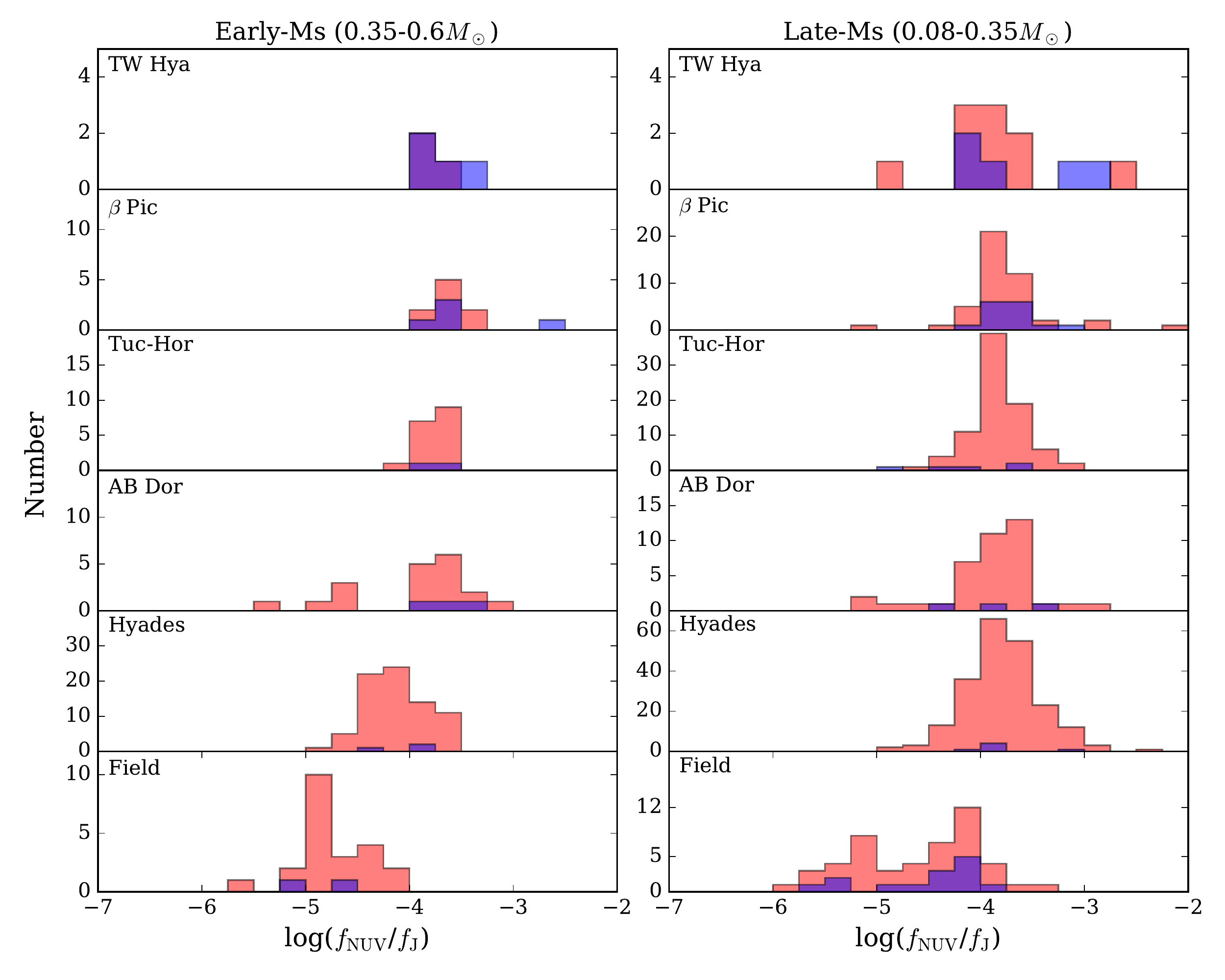}
\caption{Histograms of the {\it GALEX} NUV fractional flux densities for stars without evidence of binarity (red) and known binaries (blue), where overlapping areas appear purple. }  
\end{figure*}

\subsection{The Relationship Between {\it GALEX} FUV and NUV For Low-Mass Stars} 
The FUV to NUV ratio has been suggested to be a major factor affecting the photochemistry of the atmospheres of exoplanets around low-mass stars (e.g., \citealt{dom14}, \citealt{tian14}, \citealt{har15}).  This ratio has been show to be more than two orders of magnitude from solar-type dwarfs to M dwarfs \citep{france13}.  This change can have a significant influence on the photochemistry of planets around low-mass stars, notably on the abundances of molecules including CO$_2$, O$_2$, and O$_3$.  FUV radiation will photolyze CO$_2$, producing atomic oxygen which can then combine to form O$_2$ and eventually O$_3$.  NUV radiation on the other hand will dissociate O$_3$, and thus O$_3$ abundances are critically dependent on $f_{\rm FUV}$/$f_{\rm NUV}$ values.  Therefore, the suggestion of O$_3$ as a potential biomarker (e.g., \citealt{kalt07}) may not be appropriate for M dwarfs.  While the $f_{\rm FUV}$/$f_{\rm NUV}$ ratio has been investigated for several field age stars (\citealt{france13}, \citealt{france16}), how the $f_{\rm FUV}$/$f_{\rm NUV}$ ratio changes as an M star evolves has not yet been probed. 

Our sample of {\it GALEX}-detected M stars allows us to investigate how the FUV/NUV ratio changes as a function of both mass and age.  Note that while the FUV usually refers to the wavelength region extending from 1100--1800\AA\, which includes the most prominent UV emission line (Lyman-$\alpha$), the {\it GALEX} FUV band extends from $\sim$1350--1800\AA, and thus does not include Lyman-$\alpha$.  However, there are still a plethora of strong emission lines that originate in the chromosphere and corona of low-mass stars in the {\it GALEX} FUV passband, such as Si IV ($\lambda$1394\AA), the C IV doublet ($\lambda$1548\AA\ and $\lambda$1551\AA), and He II ($\lambda$1640\AA), and thus the {\it GALEX} FUV passband still probes this critical temperature region.  We adopt the same nomenclature for the {\it GALEX} FUV/NUV flux density ratio as \cite{miles17} ([$f_{\rm FUV}$/$f_{\rm NUV}$]$_{\rm G}$).  Figure 11 shows the [$f_{\rm FUV}$/$f_{\rm NUV}$]$_{\rm G}$ values for young stars (TW Hya, $\beta$ Pic, and Tuc-Hor), AB Dor members, and field stars as a function of mass.  We do not include the Hyades because very few Hyades members were detected in the {\it GALEX} FUV band (see Table 3).   

As with $f_{\rm NUV}$/$f_{\rm J}$ and $f_{\rm FUV}$/$f_{\rm J}$ values, we perform a survival analysis using the Kaplan-Meier estimator to account for the upper limits contained in our dataset.  Similar to previous studies (\citealt{france16}, \citealt{miles17}), we find an increase in [$f_{\rm FUV}$/$f_{\rm NUV}$]$_{\rm G}$ values as stellar mass decreases for field age M stars.  However, we see a clear distinction, by almost an order of magnitude, between young and old early M-type stars.  This age difference becomes less apparent as stellar mass decreases, and the two populations are indistinguishable for masses less than $\sim$0.3 $M_\odot$.  Thus the [$f_{\rm FUV}$/$f_{\rm NUV}$]$_{\rm G}$ flux density ratio for mid- and late-type M stars remains relatively constant over their lifetimes, while this ratio clearly evolves to lower values with age for early-type Ms.  These larger [$f_{\rm FUV}$/$f_{\rm NUV}$]$_{\rm G}$ values for young, early-Ms may have significant consequences on molecular abundances in exoplanet atmospheres at young ages.  These different evolutionary patterns should be taken into account when modeling the atmospheres of planets orbiting such stars.

\begin{figure*}
\epsscale{0.8}
\plotone{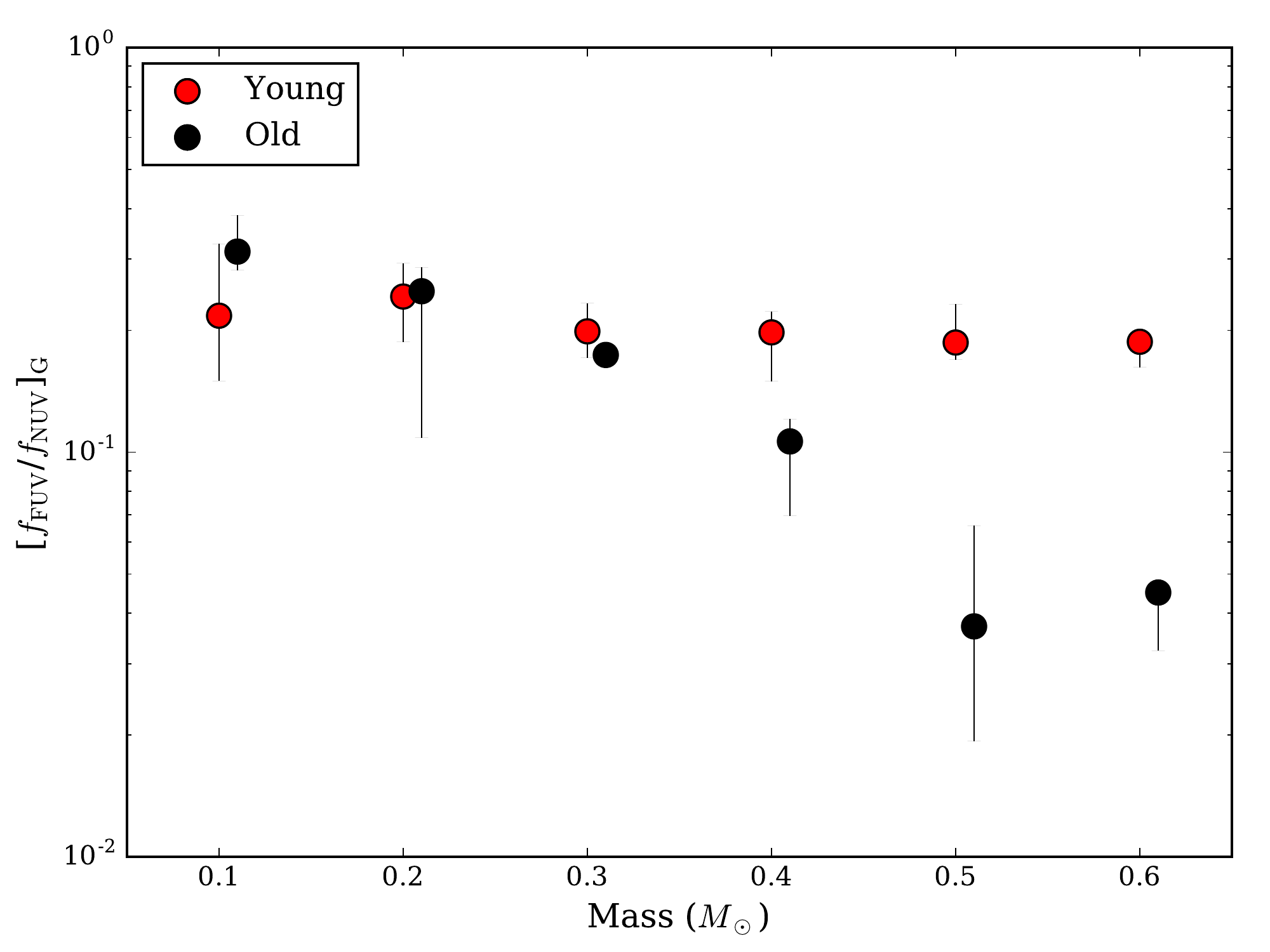}
\caption{The median {\it GALEX} FUV/NUV fractional flux density for young M stars (TW Hya, $\beta$ Pic, and Tuc-Hor) and field-age M stars as a function of mass.  Some symbols are slightly offset along the abscissa for differentiation purposes.  Error bars represent inner quartiles. }  
\end{figure*}

The origin of the different [$f_{\rm FUV}$/$f_{\rm NUV}$]$_{\rm G}$ evolutionary patterns between early- and late-Ms can be deduced by comparing the $f_{\rm NUV}$/$f_{\rm J}$ and $f_{\rm FUV}$/$f_{\rm J}$ ratios discussed in Section 4.1 and seen in Figures 3 through 9.  The $f_{\rm NUV}$/$f_{\rm J}$ and $f_{\rm FUV}$/$f_{\rm J}$ ratios for late-type Ms decrease by a similar amount from young to field ages, resulting in relatively constant  [$f_{\rm FUV}$/$f_{\rm NUV}$]$_{\rm G}$ ratios across all ages.  For early Ms, the $f_{\rm FUV}$/$f_{\rm J}$ ratio has a steeper decrease from young to field ages compared to the $f_{\rm NUV}$/$f_{\rm J}$ ratio, resulting in an evolving  [$f_{\rm FUV}$/$f_{\rm NUV}$]$_{\rm G}$ ratio where the contribution to the overall UV from the FUV band becomes less significant as a star ages. 

The UV evolutionary difference between young and field-age Ms can also be clearly seen when comparing absolute FUV and NUV flux densities.  Using distances for our field sample from \cite{kirk12} and for young Ms ($<$ 50 My; TW Hya, $\beta$ Pic, and Tuc-Hor) from \cite{kraus14}, \cite{bell15}, \cite{gagne17}, and \cite{shk17}, we can compare the absolute flux densities for all objects in our sample detected in both the FUV and NUV bands.  Figure 12 shows a comparison of absolute flux densities for both young and field-age Ms for objects in our sample detected in both the FUV and NUV bands.  While Ms of all masses follow a single trend for young low-mass stars, there is a clear offset between higher and lower-mass Ms in the field sample.  We perform a least-squares fit to the data using a power law of the following form to find the following relations between $f_{\rm FUV}$ and $f_{\rm NUV}$ for young Ms:     

\begin{equation}
f_{\rm FUV} = 0.56\pm0.10(f_{\rm NUV})^{0.88\pm0.03} 
\end{equation}

\noindent the following two relations for higher-mass ($>$ 0.35 $M_{\odot}$) and lower-mass ($\leq$ 0.35 $M_{\odot}$) field-age Ms:

\begin{equation}
f_{\rm FUV} = 0.09\pm0.03(f_{\rm NUV})^{0.96\pm0.10}, (M > 0.35\ M_{\odot}) 
\end{equation}

\begin{equation}
f_{\rm FUV} = 0.40\pm0.08(f_{\rm NUV})^{0.93\pm0.05}, (M \leq 0.35\ M_{\odot}) 
\end{equation}

\noindent where $f_{\rm FUV}$ and $f_{\rm NUV}$ are flux densities at 10 pc in $\mu$Jy.  Uncertainties are determined in a Monte Carlo fashion. Note that the power law slopes are all within 1$\sigma$ combined, while the offsets are significantly different.  A hint of this trend was also seen in \cite{miles17}.  These values can be used for estimating $f_{\rm FUV}$ and $f_{\rm NUV}$ values for objects for which only one or the other is available.  

\begin{figure*}
\plotone{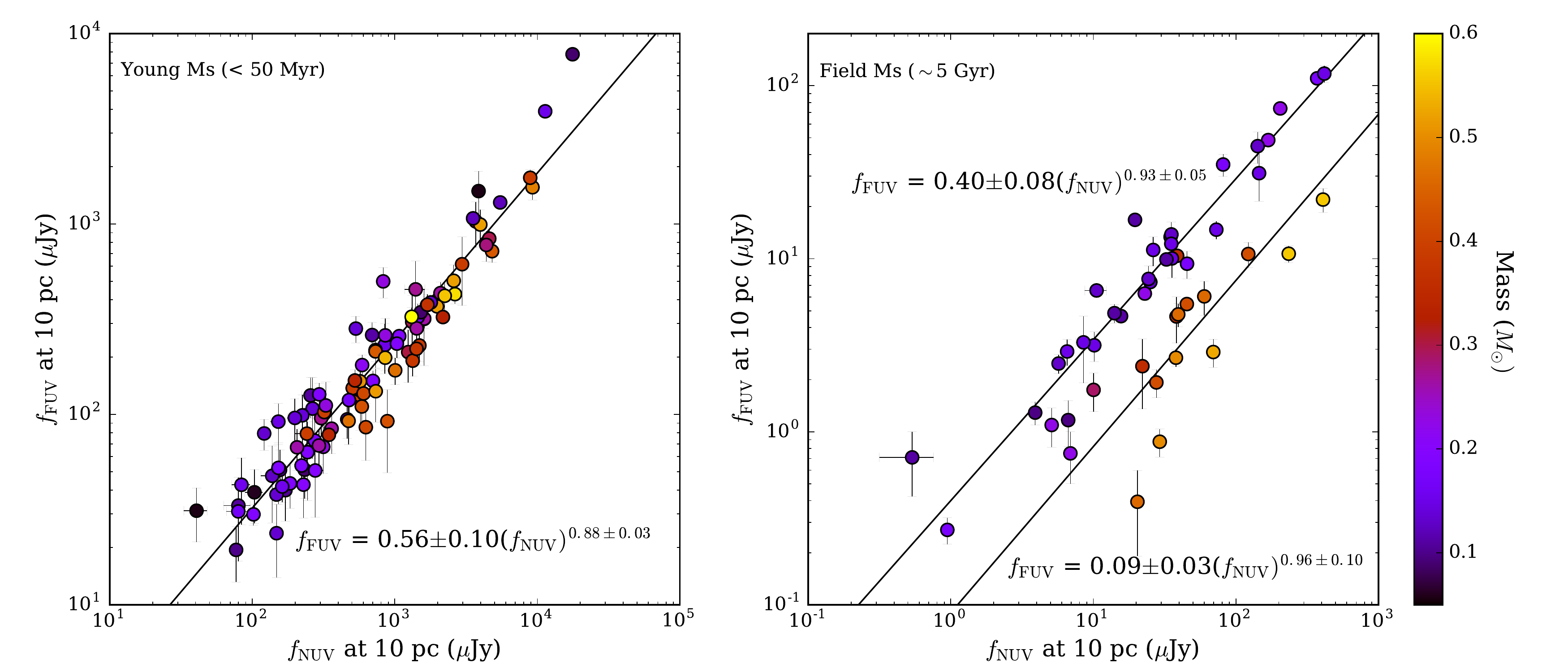}
\caption{Absolute FUV flux density versus absolute NUV flux density for young ($<$ 50 Myr; left) and field-age ($\sim$5 Gyr; right) low-mass stars.  Black-lines represent least-squares fits to the data.  For the field-age sample, we determine separate fits for higher mass ($>$ 0.35 $M_{\odot}$) and lower-mass ($\leq$ 0.35 $M{\odot}$) Ms.}  
\end{figure*}

\subsection{The Photospheric Contribution to FUV and NUV Flux Densities} 
One of the primary motivations of this work is to empirically inform low-mass stellar models with measured FUV and NUV flux densities, as stellar models for M-type stars do not include contributions from chromospheres, transition regions, or coronae, though progress is being made to include these components (\citealt{pea15}, \citealt{font16}).  Using existing low-mass stellar evolution models, we can determine how much emission in the {\it GALEX} FUV and NUV bands is due solely to the photosphere of the star. 

We investigate the photospheric contribution as a function of stellar mass and stellar effective temperature using the PHOENIX stellar atmosphere models (\citealt{hau97}, \citealt{short05}), the field-age 8 pc sample, and young stars from TW Hya, $\beta$ Pic, and Tuc-Hor with measured distances.  Distances for the 8 pc sample come from \cite{kirk12}, while distances for young moving group members come from  \cite{kraus14}, \cite{bell15}, \cite{gagne17}, and \cite{shk17}.  

Figure 13 shows the fraction of photospheric flux density to the total observed FUV and NUV flux densities as measured by {\it GALEX} as a function of both stellar effective temperature and stellar mass.  As in \cite{shk14}, we find a clear trend with $T_{\rm eff}$ in the FUV for stars between 3400 K and 3800 K, where the photospheric contribution becomes less significant with decreasing $T_{\rm eff}$.  The inclusion of lower-mass stars from our sample shows that this trend continues for both old and young stars down to at least temperatures of $\sim$2800 K, though the distinction between the young and old samples becomes much less clear as $T_{\rm eff}$ decreases.  For the NUV, we again find agreement with \cite{shk14} in the 3400 K and 3800 K range, where the photospheric contribution is relatively constant for both young and old stars.  Below $\sim$3200 K, a range not probed in \cite{shk14}, we find that the photospheric contribution drops considerably for young and old stars, with the two population again becoming indistinct as $T_{\rm eff}$ decreases.  Field age Ms in our sample with $T_{\rm eff}$ $\geq$ 3200 K have an average NUV photospheric contribution of $\sim$26\%, while field age Ms with $T_{\rm eff}$ $<$ 3200 K have an average photospheric contribution of only 0.7\%.  For the young sample, there is an average $\sim$2\% photospheric contribution for $T_{\rm eff}$ $\geq$ 3200 K and 0.7\% for $T_{\rm eff}$ $<$ 3200 K. Thus the contribution to the overall flux density in the {\it GALEX} FUV and NUV bands from the stellar photosphere becomes even less significant for lower-mass M-type stars.

\begin{figure*}
\plotone{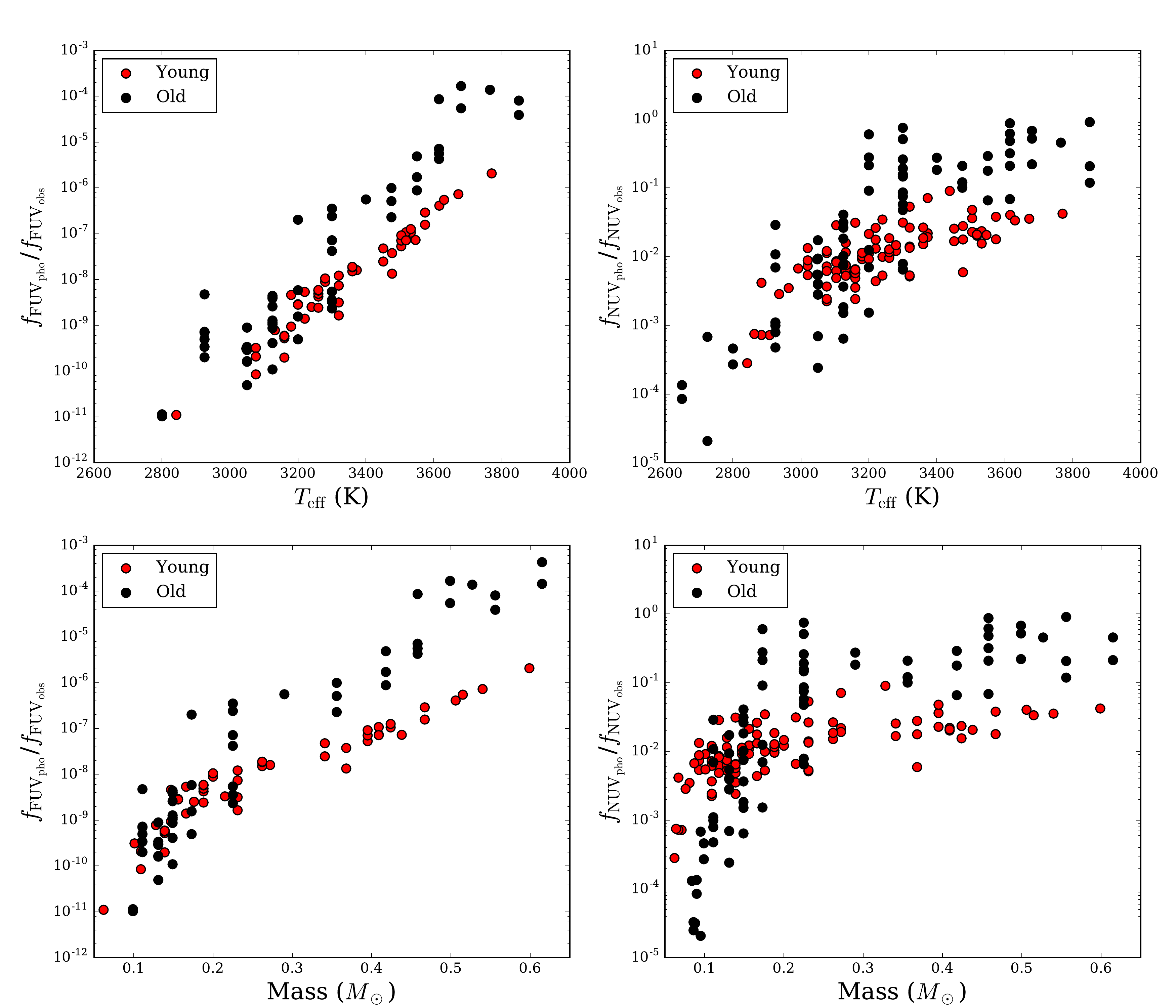}
\caption{The fraction of observed FUV (left) and NUV (right) flux density from the stellar photosphere as a function of stellar effective temperature (top) and stellar mass (bottom).}  
\end{figure*}

\section{Conclusions}

Because UV emission can have severe effects on the compositions, and existence at all, of exoplanet atmospheres orbiting low-mass stars, it is critical to know not only the current state of UV emission from planet hosts, but how emission evolved to its current state.  Using {\it GALEX} photometry and a large sample of nearby, low-mass stars with know ages, we have shown that early-Ms and mid- to late-Ms evolve very differently.  Mid- to late-Ms remain relatively active throughout their lifetimes, with only a small UV flux density decrease from young to old ages, whereas early-Ms have a much more significant decrease over the same age range.

{\it GALEX} photometry has been shown to correlate with emission from the strongest FUV emission line, Lyman-$\alpha$ \citep{shk14b}, which dominates the UV spectrum of low-mass stars (e.g., \citealt{loyd16}) and controls the photodissociation of molecules such as O$_2$, H$_2$O, CH$_4$ (e.g., \citealt{lin13}).  Furthermore, FUV flux has been shown to be strongly correlated to X-ray and EUV flux (\citealt{shk14}, \citealt{mitra05}, \citealt{france16}), which controls much of the upper atmospheric heating of orbiting planets (e.g., \citealt{lamm07}, \citealt{tian09}).  Thus, the prolonged UV activity seen for low-mass Ms, which we attribute to different rotation rates for early-Ms and fully convective mid- to late-Ms, can have serious consequences for the potential habitability and the detection of such habitability of Proxima Cen b \citep{ang16} and LHS 1140b \citep{ditt17}, and the planets around TRAPPIST-1 (\citealt{gill16}, \citealt{gill17}), which orbit M5.5, M4.5, and M8 stars, respectively.

\acknowledgments
A.\ S. and E.\ S. appreciate support from NASA/Habitable Worlds grant NNX16AB62G (PI E.\ Shkolnik). The results reported herein benefited from collaborations and/or information exchange within NASA's Nexus for Exoplanet System Science (NExSS) research coordination network sponsored by NASA's Science Mission Directorate and the NExSS grant NNX15AD53G (PI S.\ Desch). We wish to thank the anonymous referee for a helpful report that improved the quality of this work.  We thank Victoria Meadows, Travis Barman, David Charbonneau, Zach Berta-Thompson, Taisiya Kopytova, Kim Ward-Duong, and Jennifer Patience for useful discussions.  This work is based on observations made with the NASA Galaxy Evolution Explorer. {\it GALEX} is operated for NASA by the California Institute of Technology under NASA contract NAS5-98034.  This publication makes use of data products from the Two Micron All Sky Survey, which is a joint project of the University of Massachusetts and the Infrared Processing and Analysis Center/California Institute of Technology, funded by the National Aeronautics and Space Administration and the National Science Foundation.  This research has made use of the SIMBAD database, operated at CDS, Strasbourg, France. This research has made use of the Washington Double Star Catalog maintained at the U.S. Naval Observatory.

\begin{longrotatetable}
\begin{deluxetable*}{lccccccccccc}
\tabletypesize{\footnotesize}
\tablecaption{Target Sample {\it GALEX} Photometry}
\tablehead{
2MASS Designation & Disc.\tablenotemark{a} & Spec. & $J$ & FUV\tablenotemark{b} & NUV\tablenotemark{b} & $f_{\rm FUV}$/$f_{\rm J}$ & $f_{\rm NUV}$/$f_{\rm J}$ & Binary? & Binary  \\
 & Ref & Type & (mag) & (mag) & (mag) & & & & Ref}
\startdata
\colrule
\sidehead{TW Hya}
\colrule
10120908$-$3124451 & 1 & M4.0 & \phn8.848 $\pm$ 0.018 & 21.646 $\pm$ 0.318 & 20.234 $\pm$ 0.122 & \phn1.73e-05 & \phn6.36e-05 & VB (1\farcs0) & 25 \\
10284580$-$2830374 & 2 & M5.0 & 10.953 $\pm$ 0.027 & $>$22.577 & 22.325 $\pm$ 0.397 & $<$5.10e-05 & \phn6.44e-05 \\
11091380$-$3001398 & 3 & M1.5 & \phn7.629 $\pm$ 0.030 & 20.099 $\pm$ 0.138 & 18.037 $\pm$ 0.035 & \phn2.34e-05 & \phn1.56e-04 & VB (0\farcs6) & 7 \\
11102788$-$3731520 & 3 & M4.0 & \phn7.651 $\pm$ 0.019 & 17.720 $\pm$ 0.072 & 16.563 $\pm$ 0.024 & \phn2.14e-04 & \phn6.21e-04 & SB, VB (1\farcs4) & 3, 57 \\
11210549$-$3845163 & 4 & M2.0 & \phn8.999 $\pm$ 0.034 & 20.962 $\pm$ 0.425 & 19.255 $\pm$ 0.096 & \phn3.74e-05 & \phn1.80e-04 \\
11211723$-$3446454 & 4 & M1.0 & \phn8.431 $\pm$ 0.043 & 19.643 $\pm$ 0.153 & 17.711 $\pm$ 0.043 & \phn7.46e-05 & \phn4.42e-04 & VB (5\farcs0) & 58 \\
11315526$-$3436272 & 3 & M2.0 & \phn7.669 $\pm$ 0.026 & 19.213 $\pm$ 0.111 & 17.446 $\pm$ 0.009 & \phn5.50e-05 & \phn2.80e-04 & VB (0\farcs1, 2\farcs0) & 59, 60 \\
11321822$-$3018316 & 5 & M4.0 & 15.350 $\pm$ 0.047 & $>$21.780 & $>$22.819 & $<$6.10e-03 & $<$2.34e-03 & VB (80\farcs2) & 5 \\
\enddata
\end{deluxetable*}
\end{longrotatetable}

\begin{appendix}
\section{Converting {\it GALEX} Flux Densities to Fluxes}
If one wishes to convert flux density to flux (in erg cm$^{-2}$ s$^{-1}$), one can use the following formula:

\begin{equation}
f_{\rm NUV, FUV} = (f_{\rm NUV, FUV})  c \lambda_{\rm eff}^{-2} \Delta\lambda
\end{equation}
\\
\noindent where $f_{\rm NUV, FUV}$ is the flux density (in $\mu$Jy), $c$ is the speed of light, $\lambda_{\rm eff}$ is the effective filter wavelength, and $\Delta\lambda$ is the filter width.  {\it GALEX} documentation gives $\lambda_{\rm eff}$ =  2267 \AA\ and $\Delta\lambda$ = 732 \AA\ for {\it GALEX} NUV, and $\lambda_{\rm eff}$ =  1516 \AA\ and $\Delta\lambda$ =268 \AA\ for {\it GALEX} FUV\footnote{http://galex.stsci.edu/gr6/?page=faq}.   Note that the flux emitted from low-mass stars in these bands is primarily from spectral lines that originate in the chromosphere and transition region of their atmospheres.   For the NUV band in particular, much of the flux comes from Fe II lines that occur in the red side of the detector bandpass.  The amount of flux from these lines, which varies for each star, could shift the $\lambda_{\rm eff}$, affecting the resulting flux measurements.  

In order to determine the effects of M dwarf spectral shapes on $\lambda_{\rm eff}$ values, we gathered UV spectra of M dwarfs from the MUSCLES {\it HST} program (\citealt{france13}, \citealt{france16}). Using equation A2 of \cite{tok05} and the {\it GALEX} filter response curves\footnote{https://asd.gsfc.nasa.gov/archive/galex/Documents/PostLaunchResponseCurveData.html}, we calculate $\lambda_{\rm eff}$ values for seven M dwarfs with spectral types ranging from M1.5 to M5.5.  We exclude the M4.5 dwarf GJ 1214 because of the low S/N of its spectrum.  The results are shown in Table A1 and Figure A1.  Though the sample size is small, $\lambda_{\rm eff}$ values tend to decrease with later spectral type.  While the FUV $\lambda_{\rm eff}$ values bracket the {\it GALEX} provided $\lambda_{\rm eff}$ value, the calculated $\lambda_{\rm eff}$ values for the NUV band are clearly distinct from the {\it GALEX} provided value.  We find average $\lambda_{\rm eff}$ values for the FUV and NUV bands of 1542.8 \AA\ and 2553.4 \AA, respectively.   

\begin{deluxetable*}{ccccccccc}[b!]
\tablecaption{$\lambda_{\rm eff}$ for MUSCLES M Dwarfs and {\it GALEX} Filters}
\tablehead{
Star Name & Spectral & FUV & NUV  \\
 & Type & (${\rm \AA}$) & (${\rm \AA}$) }
\startdata
GJ 667C & M1.5 & 1605.2 & 2543.9  \\
GJ 832 & M1.5 & 1569.4 & 2612.9  \\
GJ 176 & M2.5 & 1553.6 & 2583.3  \\
GJ 436 & M3.5 & 1494.9 & 2551.2  \\
GJ 581 & M5 & 1503.6 & 2602.1  \\
GJ 876 & M5 & 1518.9 & 2528.2  \\
GJ 551 & M5.5 & 1553.9 & 2452.2  \\
\enddata
\end{deluxetable*}       

\begin{figure*}
\plotone{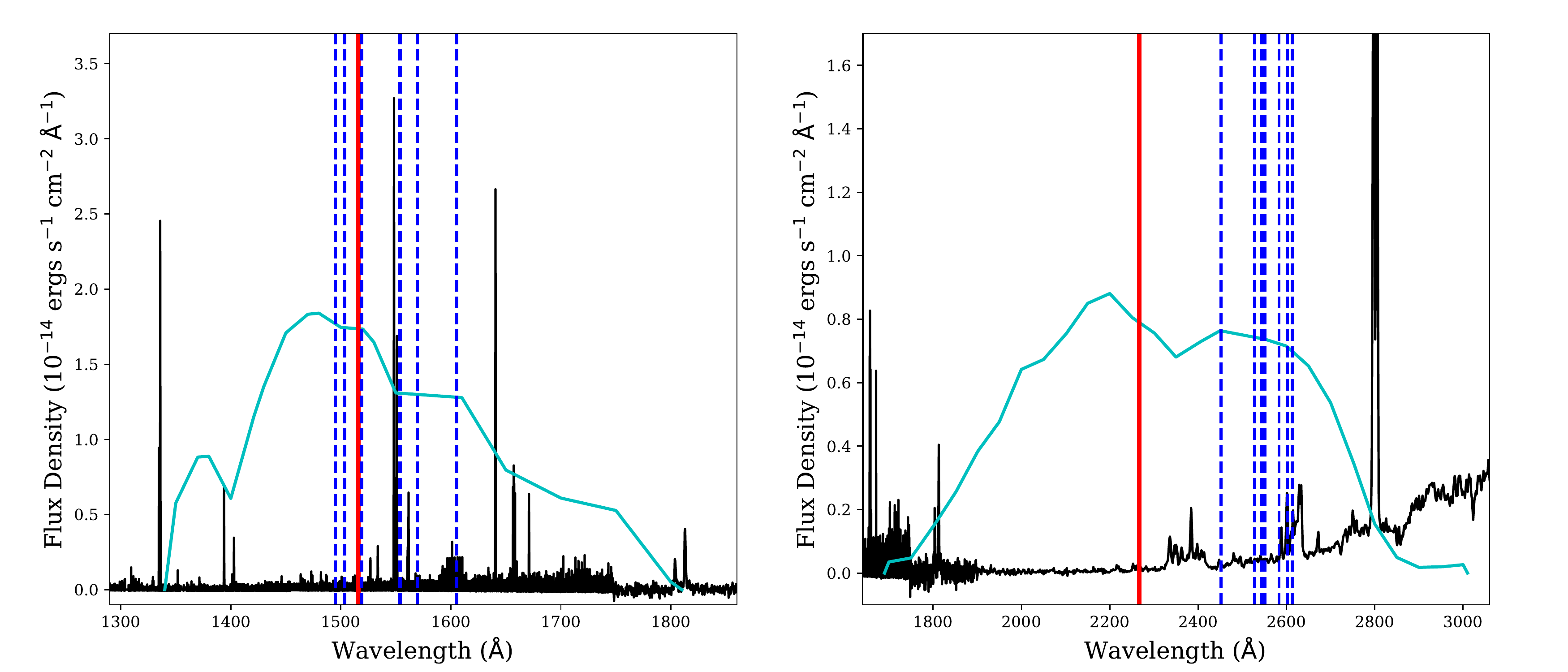}
\caption{Filter profiles of the {\it GALEX} FUV (left) and NUV (right) bandpasses (arbitrarily scaled).  The spectrum of the M2.5 dwarf GJ 176 from the MUSCLES program (\citealt{france13}, \citealt{france16}) is overplotted for comparison purposes.  The effective wavelength provided in the {\it GALEX} documentation is show as a solid red line, while the effective wavelengths determined for M dwarfs in the MUSCLES sample are show as blue dashed lines. }  
\end{figure*}

\end{appendix}

\end{document}